\documentclass[12pt]{spieman}  

\usepackage[T1]{fontenc}
\usepackage[utf8]{inputenc}
\usepackage{xspace}
\usepackage{graphicx}
\usepackage[separate-uncertainty=true,range-phrase=--,range-units=single,product-units=power,group-separator={,}]{siunitx}[=v2]
\usepackage[super]{nth}
\usepackage{amsmath}
\usepackage{amssymb}
\usepackage{booktabs}
\usepackage{xcolor}

\DeclareSIUnit\curie{Ci}

\newcommand{\ascent}{\textit{ASCENT}\xspace}
\newcommand{\xcalibur}{\textit{X-Calibur}\xspace}
\newcommand{\xlcalibur}{\textit{XL-Calibur}\xspace}

\newcommand{\sledgehammer}{\textit{SLEDGEHAMMER}\xspace}
\newcommand{\nustar}{\textit{NuSTAR}\xspace}

\newcommand{\hexp}{\textit{HEX-P}\xspace}
\newcommand{\cosi}{\textit{COSI}\xspace}

\newcommand{\ti}{$^{44}$Ti\xspace}
\newcommand{\nickel}{$^{56}$Ni\xspace}
\newcommand{\fe}{$^{56}$Fe\xspace}

\newcommand{\casA}{Cas~A\xspace}

\newcommand{\eg}{\textit{e.g.,}\xspace}
\newcommand{\ie}{\textit{i.e.,}\xspace}

\usepackage{caption}
\definecolor{dark-red}{RGB}{238,0,3}
\definecolor{dark-blue}{RGB}{9,71,171}
\definecolor{dark-green}{RGB}{17,149,17}
\definecolor{deep-purple}{RGB}{165,0,220}
\definecolor{dark-yellow}{RGB}{175,175,0}

\title{ASCENT -- A balloon-borne hard X-ray imaging spectroscopy telescope using transition edge sensor microcalorimeter detectors}

\author[a]{F. Kislat}
\author[b]{D. Becker}
\author[c]{D. Bennett}
\author[a]{A. Dasgupta}
\author[c]{J. Fowler}
\author[d]{C. Fryer}
\author[b]{J. Gard}
\author[e]{E. Gau}
\author[f]{D. Gurgew}
\author[g]{K. Harmon}
\author[f]{T. Hayashi}
\author[g]{S. Heatwole}
\author[e]{M.A. Hossen}
\author[e,h,i]{H. Krawczynski}
\author[g]{R.J. Lanzi}
\author[a]{J. Legere}
\author[c]{J.A.B. Mates}
\author[a]{M. McConnell}
\author[e,h,i]{J. Nagy}
\author[f]{T. Okajima}
\author[j]{T. Sato}
\author[c]{D. Schmidt}
\author[a]{S. Spooner}
\author[c]{D. Swetz}
\author[f]{K. Tamura}
\author[c]{J. Ullom}
\author[b]{J. Weber}
\author[a]{A. Wester}
\author[k]{P. Young}

\affil[a]{University of New Hampshire, Department of Physics \& Astronomy and Space Science Center, 8 College Rd., Durham, NH 03824, USA}
\affil[b]{University of Colorado, Department of Physics, 2000 Colorado Ave, Boulder, CO 80309, USA}
\affil[c]{NIST Boulder Laboratories, 325 Broadway, Boulder, CO 80305, USA}
\affil[d]{Los Alamos National Laboratory, Los Alamos, NM 87545, USA}
\affil[e]{Washington University in St. Louis, Physics Department, 1 Brookings Dr., CB 1105, St. Louis, MO 63130, USA}
\affil[f]{NASA’s Goddard Space Flight Center, 8800 Greenbelt Rd, Greenbelt, MD 20771, USA}
\affil[g]{NASA Wallops Flight Facility, 32400 Fulton St., Wallops Island, VA 23337, USA}
\affil[h]{McDonnell Center for the Space Sciences at Washington University in St. Louis}
\affil[i]{Quantum Sensor Center at Washington University in St. Louis}
\affil[j]{Rikkyo University, 3-34-1 Nishi Ikebukuro, Toshima-ku, Tokyo 171-8501, Japan}
\affil[k]{Arizona State University, School of Earth and Space Exploration,
Tempe, AZ 85287, USA}

\begin{document} 
\maketitle

\begin{abstract}
Core collapse supernovae are thought to be one of the main sources in the galaxy of elements heavier than iron.
Understanding the origin of the elements is thus tightly linked to our understanding of the explosion mechanism of supernovae and supernova nucleosynthesis.
X-ray and gamma-ray observations of young supernova remnants, combined with improved theoretical modeling, have resulted in enormous improvements in our knowledge of these events.
The isotope \ti is one of the most sensitive probes of the innermost regions of the core collapse engine, and its spatial and velocity distribution are key observables.
Hard X-ray imaging spectroscopy with the Nuclear Spectroscopic Telescope Array (\nustar) has provided new insights into the structure of the supernova remnant Cassiopeia~A (\casA), establishing the convective nature of the supernova engine.
However, many questions about the details of this engine remain.
We present here the concept for a balloon-borne follow-up mission called \ascent (A SuperConducting ENergetic x-ray Telescope).
\ascent uses transition edge sensor gamma-ray microcalorimeter detectors with a demonstrated \SI{55}{eV} Full Width Half Maximum (FWHM) energy resolution at \SI{97}{keV}.
This 8--16-fold improvement in energy resolution over \nustar will allow high resolution imaging and spectroscopy of the \ti emission.
This will allow a detailed reconstruction of gamma-ray line redshifts, widths, and shapes, allowing us to address questions such as: What is the source of the neutron star ``kicks''? What is the dominant production pathway for \ti? Is the engine of \casA unique?
\end{abstract}


{\noindent \footnotesize\textbf{*}Fabian Kislat, \linkable{fabian.kislat@unh.edu}}

\section{Introduction}
Core-collapse supernovae (CCSNe) of prior generations of stars are thought to be a major source of elements heavier than iron in our Solar System~\cite{burbidge_etal_1957}.
Hence, understanding their explosion mechanism is key to understanding the evolution of our Galaxy eventually supporting life on Earth.
In this paper, we describe a concept for a new balloon-borne high-energy X-ray telescope and potential future satellite mission that will, among other science goals, provide new experimental insights into the inner workings of the CCSN engine.

Supernovae as a source of heavy elements are supported both by theoretical considerations and experimental evidence.
Since \fe is the nucleus with the lowest mass per nucleon, fusion of heavier elements cannot serve as a source of energy in stellar cores.
Instead, heavier elements are formed via slow (\textit{s}-process) or rapid neutron capture (\textit{r}-process).
Some of the earliest and strongest experimental evidence for nucleosynthesis in CCSNe comes from the detection of \SI{847}{keV} and \SI{1238}{keV} gamma-rays associated with the decay of $^{56}\text{Co}$ to \fe in SN~1987A~\cite{matz_etal_1988}.
Direct evidence for the fact that our Solar System is indeed made of reprocessed stellar ejecta comes from the analysis of presolar grains in meteoritic material and interplanetary dust, whose isotopic composition is representative of the seed material of the Solar System.\cite{choi_etal_1998}
In fact, recent simulations show that $15M_\odot$ CCSNe are capable of producing many of the isotopic anomalies found in certain presolar SiC grains~\cite{schulte_etal_2021}, which have long been argued to condense in supernovae~\cite{amari_etal_1992}.

An important conclusion from the observations of high-energy X-ray and gamma-ray emission from SN~1987A soon after the explosion was that mixing of material from the different shells of the progenitor star must occur very early on in the explosion~\cite{herant_benz_1992}.
This mixing moves radioactive nickel outward from the innermost parts of the ejecta, which then drives the X-ray emission.
However, the details of this mixing and the underlying mechanism are still poorly understood and depend on the local conditions of the early shock, such as peak temperature and density.
Due to these convective instabilities, anisotropies are expected in supernovae and their remnants.

The structure of young supernova remnants (SNR) reflects the conditions of the explosion.
Since many galactic SNR are spatially resolvable in X-rays and gamma-rays, these remnants are an excellent site to study supernova explosion physics.
Regions where explosive Si burning occurs can be observed via the K line emission from \fe, which is a decay product of \nickel produced during Si burning with relatively little dependence on local conditions.
The observations, however, come with the caveats that the X-ray emission depends on the heating of the material in the shock, and that some of the iron may actually be interstellar material swept up in the shock rather than supernova ejecta.

The production of \ti in the same regions, on the other hand, is very sensitive to the local conditions.
This isotope with a half-life of \SI{58.9+-0.3}{yr}~\cite{ahmad_etal_2006} is in principle observable in galactic supernova remnants up to a few hundred years old.
It decays via $^{44}\text{Ti} \to {}^{44}\text{Sc} \to {}^{44}\text{Ca}$, emitting gamma-rays with energies of \SIlist{1157;78.32;67.87}{keV} with branching ratios between \SI{93}{\percent} and \SI{99.9}{\percent}~\cite{chen_etal_2011}.
These gamma-rays directly trace the distribution of \ti decays.
Because of these properties, observations of \ti are a particularly powerful tool to test supernova models, which has been noted as early as 1969~\cite{clayton_etal_1969}.

Here, we discuss a concept for a new balloon-borne high-energy X-ray telescope called \ascent (A SuperConducting ENergetic x-ray Telescope), which had been proposed to NASA's Astrophysics Pioneers program, and which could form the basis of a future \nustar follow-up mission.
\ascent consists of a novel transition edge sensor (TES) microcalorimeter gamma-ray detector array in the focal plane of a multi-layer coated Wolter-type focusing X-ray mirror.
Transition edge sensors utilize the rapid change in conductivity with temperature of a superconductor at its superconducting transition temperature $T_c$ for calorimetric energy measurements~\cite{ullom_bennett_2015}.
A gamma-ray spectrometer is constructed by coupling the TES to a thick absorbing structure, commonly made of Sn, which increases the quantum efficiency for the detection of \SIrange{10}{100}{keV} photons.
Recently, an array consisting of 512 detectors with a spectral resolution of \SI{55}{eV} FWHM at \SI{97}{keV} has been demonstrated~\cite{mates_etal_2017}, and individual detectors have achieved a resolution as precise as~\SI{22}{eV}.
Using these detectors, \ascent's spectral resolution will be about 15 times better than \nustar in the \SIrange{60}{85}{keV} energy range (\SI{900}{eV} at \SI{60}{keV}).\cite{harrison_etal_2013}

Additionally, \ascent will use a new Ni/C multilayer structure on its X-ray optics.
The energy bandpass of \nustar was limited by the platinum K edge at \SI{78.395}{keV} of its Pt/C multilayer, which prevents it from observing the blue-shifted \SI{78}{keV} \ti line.
The use of a Ni/C multilayer will extend \ascent's bandpass to \SI{85}{keV} and beyond.
Furthermore, its multilayer structure will be optimized for the \SIrange{55}{85}{keV} range, in order to maximize its effective area for observations of \ti.
While the baseline angular resolution of the \ascent optics of \ang{;2;} will be slightly worse than \nustar, it will still allow resolution of the most prominent \ti emission regions in \casA.

Observations of the supernova remnant \casA with \ascent will test if asymmetries of the ejecta can completely account for compact remnant ``kicks''.
Furthermore, they will allow us to determine the dominant pathway for the production of \ti.

So far, \ti has only been firmly detected from two objects: SN~1987A~\cite{boggs_etal_2015,grebenev_etal_2012} and \casA~\cite{iyudin_etal_1994, vink_etal_2001,grefenstette_etal_2014}.
Tentative detections from Vela~Jr.~\cite{iyudin_etal_1998} and Tycho's SNR~\cite{troja_etal_2014} have so far not been confirmed~\cite{tsygankov_etal_2016,lopez_etal_2015}.
However, the Compton Spectrometer and Imager (\cosi) will map the Galaxy with unprecedented spectral and spatial resolution and may find additional sources of \ti emission~\cite{tomsick_etal_2021}.
While \cosi is an excellent tool to discover \ti emission from additional SNR, its spatial and spectral resolution are not sufficient to map individual objects.
A more sensitive space mission based on the \ascent design could follow up on these detections and provide detailed maps of additional SNR.
This would answer the question whether features observed in \casA are universal or whether there is wide variation in the underlying engine depending on properties of the progenitor star.
Furthermore, a detection of \ti from a remnant associated with a type Ia supernova, such as Tycho's SNR, would be a major breakthrough.
Ordinarily, SNIa are not expected to produce much \ti.
However, some models predict a potential detonation below the Chandrasekhar mass limit (see \eg Woosley and Weaver, 1994~\cite{woosley_weaver_1994}), in which case a large amount of \ti may be produced. 
Such explosions are thought to be one of the candidates for the origin of Galactic positrons~\cite{crocker_etal_2017}.
Thus, such an observation would not only constrain the SNIa explosion mechanism but also provide new insights into the origin of Galactic positrons.

Launched on a stratospheric balloon from Kiruna, Sweden, \ascent will float westward at an altitude of about \SI{125000}{ft} to northern Canada.
Typical flight times are 5--7 days, allowing for up to \SI{560}{ksec} of observation time of \casA.
A Southern Hemisphere flight from McMurdo Station, Antarctica, will circle the South Pole at least once for a typical flight time of two weeks.
Such a flight will allow deep observations of SN\,1987A.

In Section~\ref{sec:objectives} we discuss the scientific questions addressed by the \ascent balloon mission, the technical aspects of which we describe in detail in Section~\ref{sec:implementation}.
In Section~\ref{sec:sensitivity}, we present results of Geant4 Monte Carlo simulations of the expected performance of \ascent and the resulting sensitivity of the instrument to address its science goals.
Finally, in Section~\ref{sec:summary}, we summarize the results and give an outlook towards a space-based mission based on the \ascent technology.

\section{Scientific objectives}\label{sec:objectives}
\ascent's improved spectral resolution will allow it to address some key questions brought up by the recent \nustar observations of \casA:  what is the source of compact remnant "kicks" and what are the conditions of \ti production in the \casA supernova remnant?

The high yield of \ti detected in \casA~\cite{iyudin_etal_1994, vink_etal_2001, renaud_etal_2006,grefenstette_etal_2014,siegert_etal_2015} is seen as strong support for the expected anisotropies~\cite{young_etal_2006}.
A \SI{2.4}{Ms} observation with the \nustar satellite was used to obtain the first 3-D map of the \ti ejecta~\cite{grefenstette_etal_2014,grefenstette_etal_2017}, which confirmed these high yields and, furthermore, found that the \ti lies in clumpy structures.
Because much of the \ti was found to lie in unshocked regions, its observation provides a pristine measurement of the asymmetries in the supernova engine, and this \nustar data decisively showed that \casA was produced from an explosion with multiple outflows (as expected from the convective engine) and not a jet. 
However, the \nustar observations left many questions unanswered and raised a series of new problems with our understanding of \casA, which \ascent will address.

Along with $^{56}$Ni, \ti is produced in the innermost supernova ejecta.
In contrast to $^{56}$Ni production, \ti production is extremely sensitive to the temperature and density evolution of the ejecta~\cite{magkotsios_etal_2010} and, hence, the nature of the explosion.
Within \numrange{1}{2} years of a supernova explosion, $^{56}$Ni decays to stable iron.
This iron produces emission lines when heated by the reverse shock as the supernova ejecta plows through the circumstellar medium and depends on the distribution of the circumstellar medium as well as the explosion, making its interpretation complicated. The innermost iron in the remnant is also difficult to measure accurately, since it is inside of the reverse shock and therefore cold.
Although attempts have been made at detecting this iron in the infrared in \casA,\cite{koo_etal_2018} the unbiased nature of the \ti observations and their sensitivity to the explosion characteristics have led to \ti's important role in shaping our understanding of the core-collapse explosion.

\subsection{The core collapse engine of \casA}%
A key to understanding the core collapse supernova engine is understanding the production and mixing of \ti during the explosion, which will identify the dominant pathways of \ti production in supernovae.
The processes that produce Ti are quasi-statistical or statistical equilibrium processes.
They produce an equilibrium distribution of nuclei based on the nuclear chemical potentials at a given temperature, density, and electron fraction~\cite{magkotsios_etal_2010}.
The local ratio between \ti and \nickel strongly constrains the thermodynamic initial conditions, allowing a fairly precise determination of the final distribution of all nuclei.
The remaining uncertainty can then be attributed to the thermodynamic history, which determines the details of freezeout.
Observations of additional nuclear species will provide additional constraints and reduce residual uncertainties.

Combining \ascent's \ti observations with observations of \casA's Si and Fe lines from JAXA's and NASA's \textsl{XRISM} mission (to be launched in 2023) will allow us to perform an improved reconstruction of the Si, Fe, and Ti configurations at the current time and at the time of the supernova explosion.
\ascent will improve the spectral resolution of the \ti emission over \nustar, allowing us to construct a more detailed map of the clumpy structures in \casA, enabling a more detailed comparison to numerical models and helping to disentangle multiple structures along a line of sight and the properties of these structures.
These comparisons provide direct constraints on the engine.

\begin{figure}
  \centering
  \includegraphics[width=.5\linewidth]{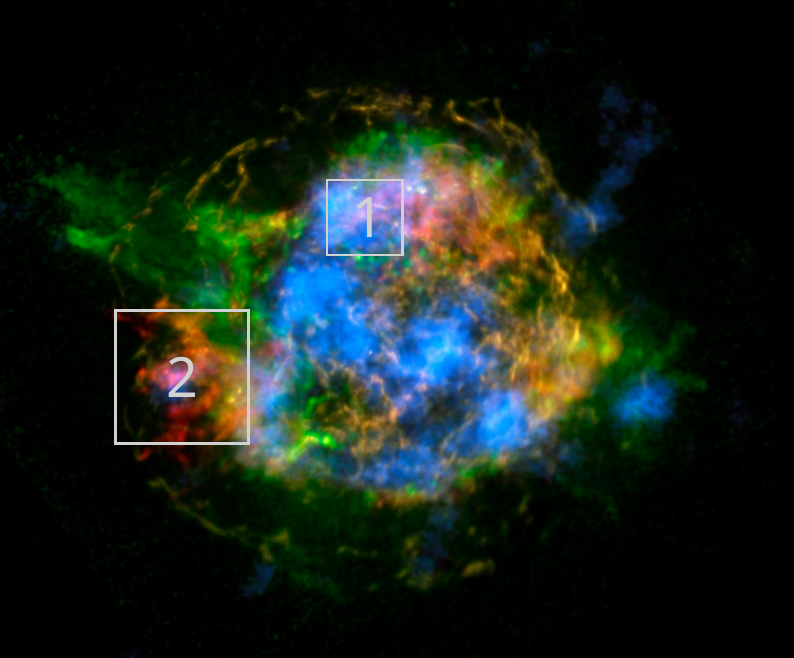}
  \caption{X-ray image of emission lines in Cassiopeia~A: iron (red), silicon/magnesium (green), titanium (blue), and continuum emission (yellow)~\cite{grefenstette_etal_2014}. Two parts of the remnant of particular interest are highlighted. \textsl{Region~1} is the only region in which \nustar detected significantly blue-shifted \ti ejecta. \textsl{Region~2} is of interest due to the highly blue-shifted almost pure iron ejecta. Composite Chandra/NuSTAR image credit NASA/JPL-Caltech/CXC/SAO.}
  \label{fig:casa}
\end{figure}
 
\nustar's results indicate that most of the ejecta are moving away from us with a velocity of \SIrange{1000}{5000}{\km\per\s} but that some \ti regions (\textsl{Region~1} in Fig.~\ref{fig:casa}, \ie Region 20 of Grefenstette et al., 2017\cite{grefenstette_etal_2017}) move towards us with \SI{7500}{\km\per\s}.
The combination of \ascent's \SI{67}{eV} energy resolution with its effective area extending to \SI{>85}{keV} will enable an analysis with much smaller systematic errors than that of the \nustar analysis.
If confirmed, these high velocity ejecta would challenge current supernova theory and provide a strong constraint on models.
Most current models that lead to a high yield of \ti do not result in velocities \SI{>4500}{\kilo\meter\per\second}~\cite{hammer_etal_2010,ono_etal_2013}, and the highest velocity found in the models by Vance et al.~\cite{vance_etal_2020} is \SI{\sim 5500}{\kilo\meter\per\second}.
The detailed velocity measurement can also be used to understand \ti production.
Nucleosynthesis calculations predict different \ti yields depending on the ejecta velocity~\cite{vance_etal_2020}.
We can use the ejecta velocities to constrain the trajectories (temperature and density evolution) of the ejecta, allowing us to test both our explosion and nucleosynthesis models.
These ejecta velocities can be tied to more fundamental properties like the electron fraction and nuclear cross sections.
  
Because we expect \ti to be produced at some level whenever $^{56}$Ni is produced, we expect to see \ti lines wherever iron is observed.
This raises the question why there are large iron ejecta with no evident detection of \ti in \casA.
The \nustar observations could not detect \ti in the iron-rich southeastern region of \casA (\textsl{Region 2} in Fig.~\ref{fig:casa}).
Is it because that iron is produced directly (not the decay product of $^{56}$Ni), or was there a large amount of $^{56}$Ni produced with \ti mass fractions below the \nustar detection limit?
The \nustar upper limit in this region of the SNR is not very constraining,\cite{grefenstette_etal_2017} and the iron-rich ejecta are blue-shifted with a velocity up to \SI{3000}{\kilo\meter\per\second}~\cite{willingale_etal_2002, delaney_etal_2010}.
Therefore, \ascent may be able to detect the \ti in this region due to its sensitivity to the blue-shifted \SI{78}{keV} line, or significantly improve on the \nustar upper limit.

\subsection{Compact Remnant Kicks}
Observations of pulsar proper motions and the existence of specific peculiar binary systems suggest that momentum is imparted onto compact remnants during their formation~(for a review, see Fryer and Kusenko, 2006\cite{2006ApJS..163..335F}).
A diverse set of models have been proposed to create these kicks, but these models can be separated into two categories:  asymmetries in the ejecta and asymmetries in the neutrino emission.
Under the convection-enhanced supernova engine paradigm~\cite{1994ApJ...435..339H}, low-mode convection produces asymmetric explosions with a nonzero net momentum in the ejecta~\cite{1995PhR...256..117H}.
These asymmetries impart a net momentum to the compact remnant, and a \SI{1}{\percent} asymmetry in the ejecta produces the high observed kick velocities.
Although simulations have struggled to produce some of the highest observed kicks, ejecta asymmetries remains one of the strongest candidate mechanisms for explaining pulsar proper motions.
Alternatively, asymmetries in the neutrino emission (typically requiring strong magnetic fields -- albeit not necessarily strong bipolar magnetic fields) also carries away a net momentum, imparting an equally strong kick onto the compact remnant.\cite{socrates_etal_2005}

The different mechanisms proposed in the ejecta and neutrino mechanisms make a variety of predictions on the relation of the compact remnant kicks with relation to angular momentum, dipole magnetic-field strength, final remnant mass, and the formation of a black hole versus a neutron star.
Many of these predictions are indirect, and it is difficult to place strong constraints on the mechanism with existing observations.
However, \nustar observations opened up the potential for a more direct observational constraint with detailed maps of the \ti to compare the asymmetries in the explosion to the remnant velocities.
Because \ti is produced in the innermost ejecta, it is an ideal probe of these explosion asymmetries.
However, to truly compare the explosion asymmetries with the compact remnant kick, we need detailed 3-dimensional ejecta information.
Although the current \nustar data hinted at a correlation between the explosion asymmetries and the compact remnant kick supporting the ejecta kick mechanism~\cite{grefenstette_etal_2017}, the higher-fidelity \ascent observations will allow a more quantitative test of the ejecta kick mechanism.

\subsection{Validating the convective SN engine}
The convective nature of the supernova engine in \casA has been established quite firmly.
However, there are only two SNR with confirmed detections of \ti, and \casA is the only SNR in which \ti emission has been spatially resolved.
This raises the question whether the SN engine of \casA is unique, which \ascent can address through observations of SN\,1987A.\cite{boggs_etal_2015}
While \ascent cannot spatially resolve SN\,1987A, a precise measurement of the \ti line shapes can be used to quantify asymmetries in the Ti distribution.

\begin{figure}
  \centering
  \includegraphics[width=.7\linewidth]{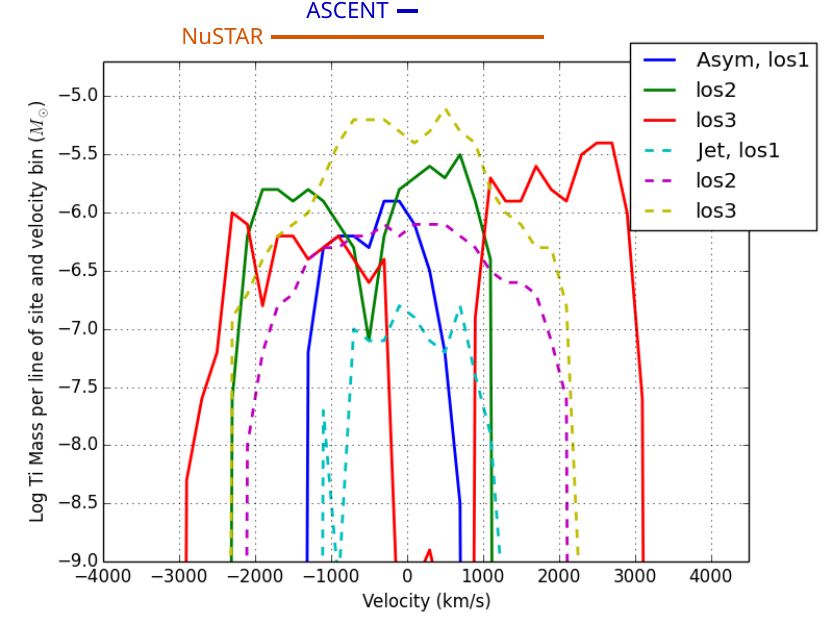}
  \caption{%
    Velocity distribution of the \ti ejecta for two different supernova explosions: a bipolar explosion where the ejecta is fastest along the axis (jet) and an explosion with multiple strong outflows mimicking the predictions of the convective supernova engine (asym).\cite{ellinger_etal_2012} The structure in the line of sight velocity distribution can be tied to the structure of the supernova engine (\ie the bipolar explosion has a very different profile than the asymmetric explosion). The bars at the top of the graph illustrate the line-of-sight velocity resolution of \nustar and \ascent. Thanks to \ascent's energy resolution, we will be able to measure these differences.%
    }
  \label{fig:losvelocity}
\end{figure}

Figure~\ref{fig:losvelocity} shows the velocity distributions along three lines of sight for two different supernova explosions.
These velocity distributions are derived from 3-dimensional smooth particle hydrodynamics simulations of asymmetrically-driven supernova explosions.\cite{ellinger_etal_2012}
One explosion is bimodal (either produced by a mild ``jet'' or low-mode convection model with rotation) and the other is more representative of a low-rotation, low-mode convectively-driven explosion (``Asym'').
For the velocity distributions in this figure, we chose 3 different lines-of-site and measured the velocities of the ejecta along these lines-of-site (to determine the red- and blue-shifted features).
The \ascent observations will not only be able to easily differentiate between these models, but also enable us to further constrain the specific features of the convective engine.

\section{Technical implementation}\label{sec:implementation}
\subsection{Overview}
\begin{figure}
    \centering
    \includegraphics[width=.75\linewidth]{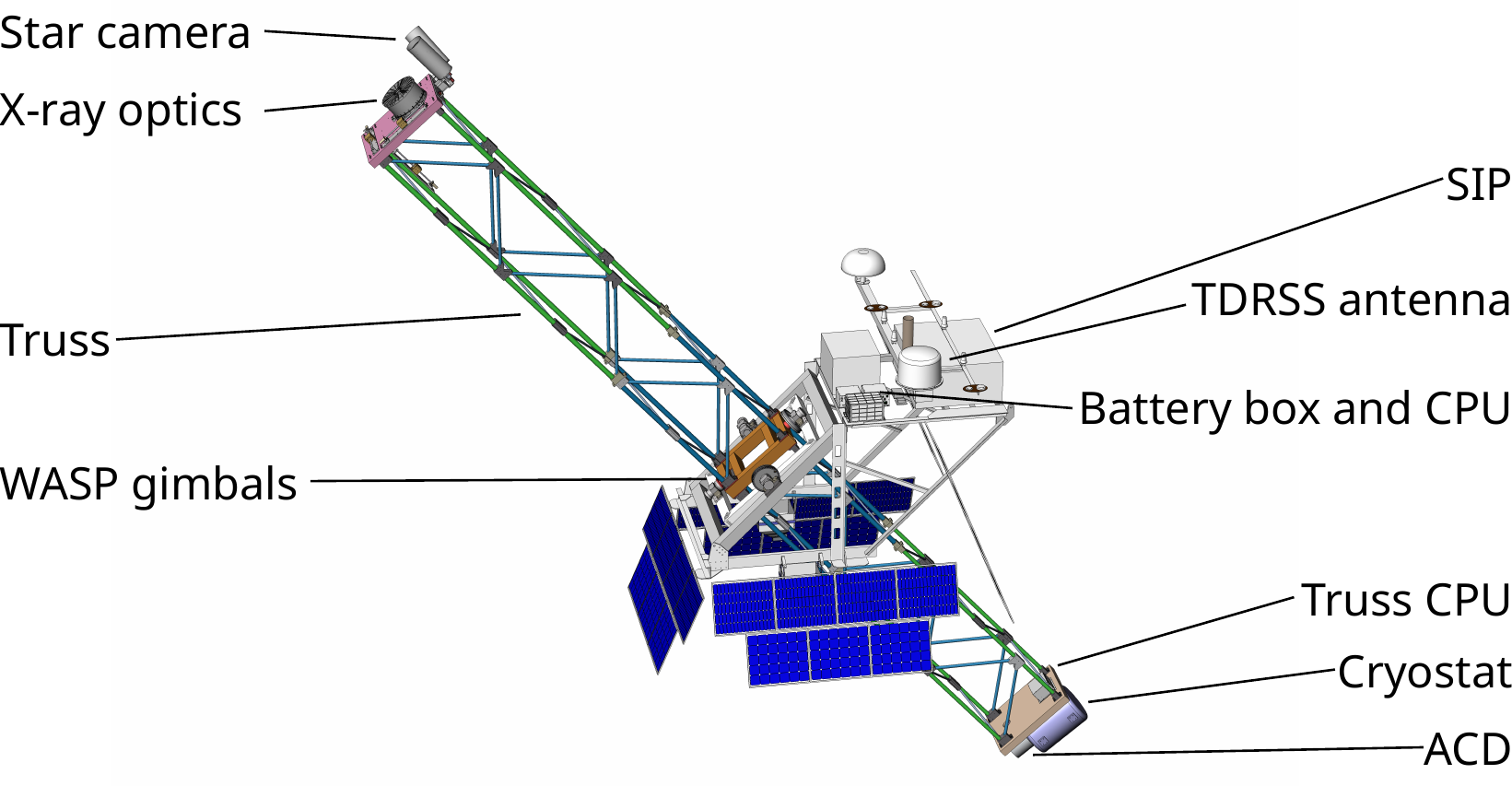}
    \caption{CAD rendering of the \ascent telescope. Key components are labeled in the Figure and described in Section~\ref{sec:implementation}. The gondola and truss design are almost identical to \xlcalibur, maximizing flight heritage.}
    \label{fig:telescope}
\end{figure}

The \ascent experiment (Figure \ref{fig:telescope}) uses a \SI{12}{m} optical bench with a \SI{45}{cm}-diameter, $F = \SI{12}{m}$ multilayer X-ray mirror at the front end and a cryogenically cooled microcalorimeter detector assembly at the rear end.
A balloon gondola holds a two-frame gimbal, pointing the optical bench in the direction of the observed astrophysical sources with the help of the Wallops Arc Second Pointer (WASP) system~\cite{stuchlik_2015}.
The microcalorimetric detector array is cooled by an Adiabatic Demagnetization Refrigerator (ADR) inside a \SI{65}{L} liquid He dewar.
Table~\ref{tab:summary} summarizes key characteristics of the \ascent observatory and its expected performance.
In the remainder of this section, we describe the design of each of the main components.

\begin{table*}
  \centering
  \caption{%
    Key \ascent payload characteristics and expected performance.
    Details of the performance estimates are provided in Section~\ref{sec:sensitivity}.%
    }
  \begin{tabular}{lp{.4\linewidth}p{.35\linewidth}}
   \toprule
    Component & Description & Performance \\
   \midrule
    Truss & Carbon fiber tubes and aluminum joints & Focal spot movement \SI{<3}{mm}, alignment knowledge \SI{0.5}{mm} (\ang{;;9}) \\
    Pointing system & Pitch-yaw articulated & Pointing precision 1.0--\ang{;;3.6} ($3\sigma$) on source \\
    Star camera & \SI{100}{mm}, f/1.5 short-wave infrared lens & Pointing knowledge ${<}\ang{;;15}$ ($3\sigma$) \\
    X-ray mirror & Wolter I, \SI{12}{m} focal length, diameter \SI{40}{cm}, 110 Ni/C-coated and 100 Pt/C-coated shells & Effective area \SI{190}{cm^2} at \SI{70}{keV}, Angular resolution \ang{;2;} HPD, Field of view \ang{;5;} FWHM \\
    Cryostat & LHe-backed adiabatic demagnetization refrigerator & Base temperature \SI{70}{mK} \\
    Detector & Two-layer gamma-ray TES array, 256 pixels each, \SI{1.4x1.4x0.59}{mm} absorbers ($\ang{;;30}\times\ang{;;30}$ at \SI{12}{m}), microwave multiplexed readout & Bandpass: \SIrange{2}{100}{keV}, energy resolution $\Delta E(\SI{80}{keV}) = \SI{67}{eV}$ FWHM \\
    Power & Detectors, cryostat, heaters & \SI{\sim 350}{W} \\
    Mass & Mass under balloon rotator & \SI{\sim 1700}{kg} \\
   \midrule
    Signal rate & 1 Crab source at \ang{45} elevation & \SI{.5}{Hz} at \SIrange{60}{80}{keV} \\
    Background rate & BGO shield veto applied & \SI{0.02}{Hz} at \SIrange{60}{80}{keV} \\
   \bottomrule
  \end{tabular}
  \label{tab:summary}
\end{table*}

The telescope will be carried by a \SI{1.1e6}{\cubic\meter} He-filled balloon to an altitude of about \SI{38}{km}.
When launched from Esrange in Kiruna, Sweden, it will partially circle the North Pole, reaching Northern Canada after a typically \numrange{5}{7}-day flight.
On this trajectory, the telescope will be able to continuously observe \casA with an elevation angle of about $36 - \ang{82}$.
Additional, longer, flights from McMurdo Station (Antarctica) will circle the South Pole, enabling deep observations SN~1987A.

\subsection{Focal plane instrumentation}
\subsubsection{Transition Edge Sensor array}
Microcalorimeter technology has shown great promise for transforming X-ray astrophysics. 
For example, the \textsl{Hitomi} mission used a 36-pixel Si thermistor microcalorimeter array for its Soft X-ray Spectrometer (SXS)~\cite{aharonian_etal_2016}.
The \textsl{Athena} mission (to be launched in 2032) will use a \num{4000}-pixel microcalorimeter array for its X-ray Integrated Field Unit (X-IFU)~\cite{smith_etal_2016,barcons_etal_2017}.
Over the last 15 years, transition-edge sensor (TES) microcalorimeter spectrometers have been developed as cutting-edge tools in the fields of nuclear materials analysis~\cite{bennett_etal_2012,winkler_etal_2015,hoover_etal_2015,mates_etal_2017} and the X-ray sciences~\cite{uhlig_etal_2013,miaja_avila_etal_2016,palosaari_etal_2016,okada_etal_2016,oneil_etal_2017,doriese_etal_2017,titus_etal_2017}.

\begin{figure*}
  \centering
  \includegraphics[width=\textwidth]{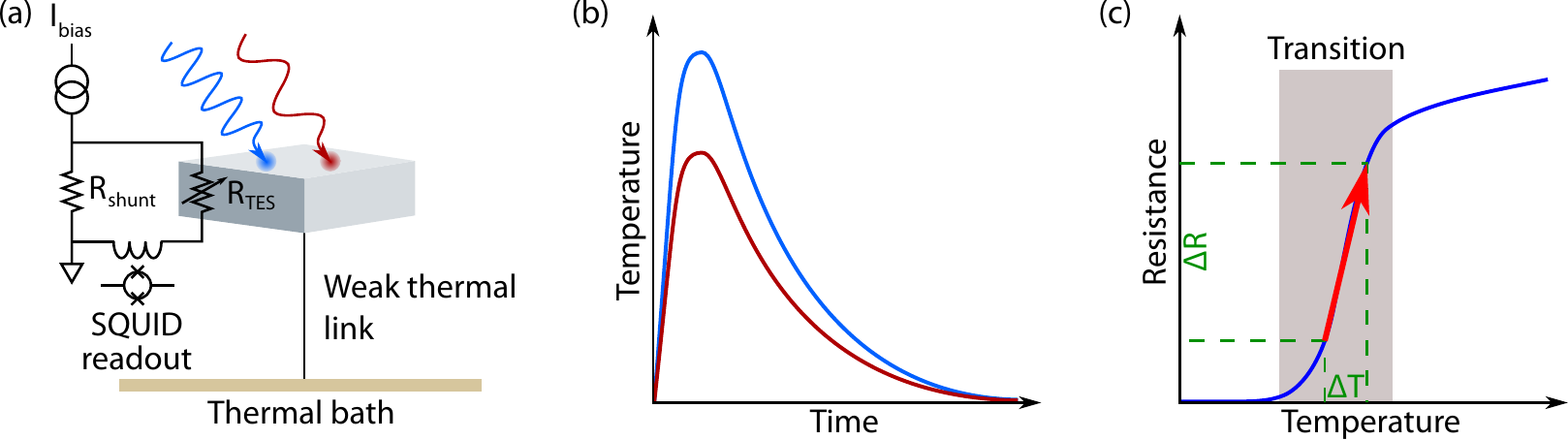}
  \caption{(a) Calorimetric spectroscopy of x-rays.
    An incident photon deposits its energy into a target with a weak thermal link to a cold isothermal bath. 
    (b) A typical pulse-response curve with a decay time determined by the properties of the calorimeter element and its coupling to the isothermal bath. The filtered pulse height is an extremely precise measure of the photon's energy.
    (c) Thermometry is performed with a thin-film superconducting transition-edge sensor.
    The extreme precision results from the sharp temperature dependence of the electrical resistance of the thin film operated close to its superconducting transition temperature.
    }
  \label{fig:tes-principle}
\end{figure*}

TES microcalorimeters are detectors that measure the energy of individual photons through the temperature change of a superconducting thin film thermometer (see Fig.~\ref{fig:tes-principle}).
The TES thermometer is coupled to a photon absorber composed of a high-Z element such as bismuth or tin, enabling high quantum efficiency for x-rays up to \SI{100}{keV}.
The fundamental energy resolution of a calorimeter is
\begin{equation}\label{eq:tes_resolution}
  \Delta E \propto \sqrt{k_B T^2 C},
\end{equation}
where $T$ and $C$ are the sensor temperature and heat capacity~\cite{irwin_hilton_2005,ullom_bennett_2015}, allowing these devices to achieve extraordinary energy resolution by operating at cryogenic temperatures of about~\SI{\sim 100}{mK}.

\begin{figure*}
  \centering
    \includegraphics[width=\textwidth]{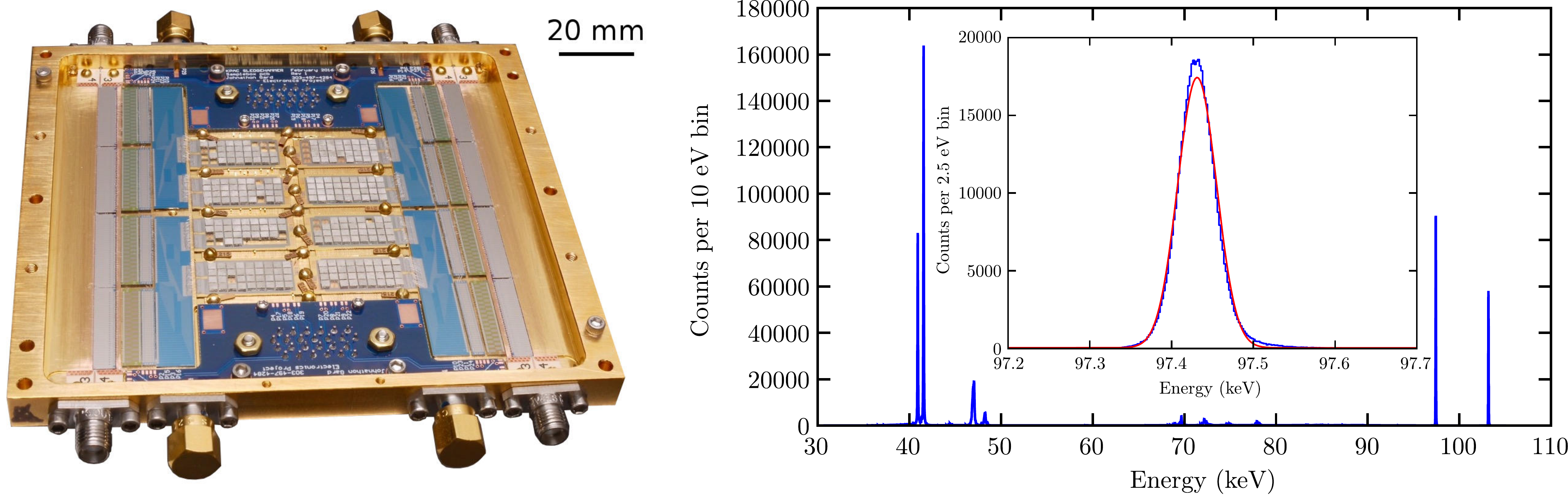}
  \caption{\emph{Left:} Photograph of the fully assembled \sledgehammer detector package.
    The package contains eight TES microcalorimeter chips with 32 sensors each (center), eight microwave multiplexer chips with 32 channel readout (outer vertical columns), and eight chips each for detector bias, Nyquist filtering, and signal routing.
    The TES signals are read out by two pairs of coaxial cables attached to the box by SMA connectors on the top and bottom of the box, each recording the signals for 128 sensors.
    \ascent will use similar architecture to minimize risk, but with two monolithic detector chips stacked on top of each other, to minimize inter-pixel dead space and maximize collection efficiency.
    \emph{Right:} A combined $^{153}$Gd spectrum from 89 active TESs measured simultaneously using microwave SQUID multiplexing readout.
    The inset shows a zoomed region around the \SI{97}{keV} $\gamma$-ray peak (blue) with a Gaussian fit FWHM resolution of \SI{55}{eV} (red). The energy resolution achievable with TES microcalorimeters is ~15 times better than that achieved by \nustar, achieving \SI{270}{km/s} accuracy in measurements of the velocity of $^{44}$Ti ejecta. Reprinted from Mates et al.\ (2017)\cite{mates_etal_2017} with the permission of AIP Publishing.
    }
  \label{fig:sledgehammer}
\end{figure*}

A prototype detector array called ``Spectrometer to Leverage Extensive Development of Gamma-ray TESs for Huge Arrays using Microwave Multiplexed Enabled Readout'' (\sledgehammer, see Fig.~\ref{fig:sledgehammer} and Mates et al., 2017\cite{mates_etal_2017}) has achieved a full-width at half maximum (FWHM) resolution of \SI{55}{eV} at \SI{97}{keV}.
A resolution as low as \SI{22}{eV} has been demonstrated with individual detectors~\cite{bacrania_etal_2009, bennett_etal_2012, mates_etal_2017}.
The \SI{100}{keV} energy resolution of \SI{55}{eV} FWHM of \sledgehammer is $10\times$ better than that of cryogenically cooled High Purity Germanium spectrometers (HPGe) and ${\sim}20\times$ better than room-temperature CdZnTe detectors.

\begin{figure*}
\centering
\includegraphics[width=.8\textwidth]{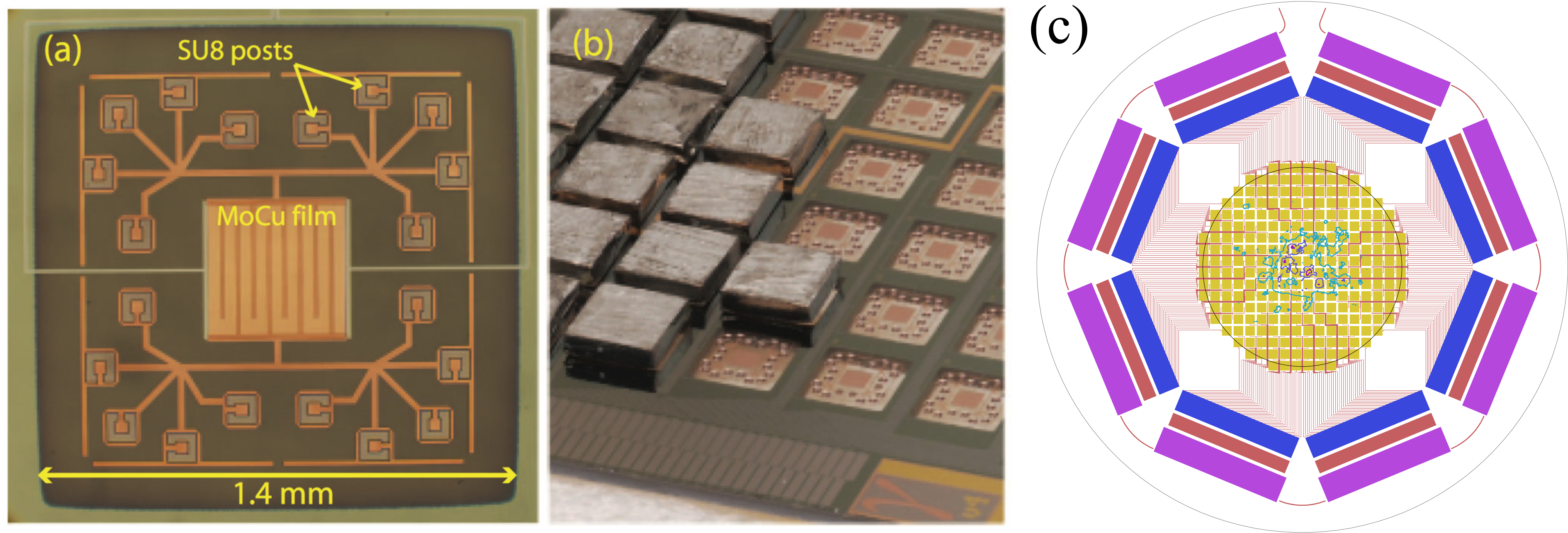}
\caption{%
  (a) Photograph of a TES gamma-ray microcalorimeter pixel before the Sn absorber is attached, showing the $\text{Si}_3\text{N}_4$ membrane (darker area), the Mo-Cu TES in the middle, and 20 SU8 epoxy posts connected to the TES by the Cu legs.
  (b) A portion of the detector chip with some of the Sn absorbers attached.
  The Sn absorbers are \SI[product-units=repeat]{1.45 x 1.45 x 0.38}{mm} thick, placed on a \SI{1.75}{mm} pitch.
  (c) Each \ascent detector die uses a central array of 256 of these $\gamma$-ray sensors (within the black inner circle) surrounded by an octagonal pattern of eight sets of 32-channel bias chips (blue), Nyquist filtering chips (red) and microwave SQUID multiplexer chips (purple).
  For scale, the outer black circle is \SI{80}{mm} in diameter, and the overlay indicates a projection of the \casA \ti distribution measured by \nustar. 
  Each die will be fabricated monolithically from a \SI{75}{mm} Si wafer to minimize space between pixels.
  Figures (a) and (b) reprinted from Bennett et al.\ (2012)\cite{bennett_etal_2012} with the permission of AIP Publishing.
  }
  \label{fig:tes-image}
\end{figure*}

\ascent will use detectors similar to those of the \sledgehammer hard X-ray/$\gamma$-ray spectrometer~\cite{mates_etal_2017}.
A photograph of a \sledgehammer detector is shown in Fig.~\ref{fig:tes-image}.
These sensors use polycrystalline tin absorbers to absorb photons.
Tin is chosen because it combines a relatively high stopping power for $\gamma$-rays in the energy range of interest with a low specific heat at cryogenic temperatures.
In \sledgehammer, these absorbers are \SI[product-units=repeat]{1.45 x 1.45}{mm} in area and \SI{0.38}{mm} thick.
For \ascent we plan on using \SI{0.59}{mm} thick absorbers to increase quantum efficiency to \SI{87}{\percent} at \SI{68}{keV} and \SI{75}{\percent} at \SI{78}{keV}.
The \SI{55}{\percent} increase in absorber volume results in a corresponding increase of the heat capacity, and a \SI{25}{\percent} increased energy resolution based on Eq.~\eqref{eq:tes_resolution}.
The expected energy resolution of the detectors, thus, increases to~\SI{68}{eV}. 
The absorbers are glued to epoxy posts, which are connected to the TES element by copper traces of equal length to ensure a uniform thermal path to the sensor (Fig.~\ref{fig:tes-image}).
The TES element is a \SI[product-units=repeat]{400 x 400}{\micro\meter} bilayer of superconducting material and normal metal, lithographically deposited on a $\text{Si}_3\text{N}_4$ membrane.
The transition temperature $T_c$ is set to ${\sim}\SI{120}{mK}$ by the superconducting proximity effect in thin-film bilayers.
This $T_c$ includes enough margin above the base temperature of an Adiabatic Demagnetization Refrigerator (ADR) to allow for stable operation.
Options for bilayer materials include MoCu, as in \sledgehammer, as well as MoAu TES using NIST's patented hasTES process.\cite{weber_2020}

Current fabrication methods require manual placement of the absorbers on the TES array using mechanical tweezers.
This constrains the minimum size and spacing of the absorbers to dimensions close to those of \sledgehammer, resulting in a \SI{1.75}{mm} pixel pitch and an array fill fraction of about~\SI{65}{\percent}.
At \SI{12}{m} focal length, this pixel pitch corresponds to an angular separation of~\ang{;;30}.
The low array fill fraction correspondingly reduces the overall photon collection efficiency of the array.

To alleviate these issues, the instrument detector package will consist of 512 detectors, in the form of two dies each containing 256 detectors.
The dies will be stacked on top of each other and offset so that the detectors in the lower die will lie directly underneath the gaps between detectors in the upper die.
This will result in a total detection efficiency for photons striking the array of \SI{80}{\percent} at \SI{68}{keV} and \SI{70}{\percent} at \SI{80}{keV}. 
The layout and design of each die is conceptually similar to the proven design of the \sledgehammer microcalorimeter array.
The central array of 256 TES detectors in each die (Fig.~\ref{fig:tes-image}(c)) is fabricated from a single \SI{75}{mm} Si wafer, with wiring to carry the TES signals to bond pads for connection to the rest of the readout circuitry arranged in an octagon around the outside of the TES array.

At \SI{12}{m} focal length, a point spread function with the half power diameter (HPD) of \ascent corresponds to \SI{3.5}{mm}, which is Nyquist sampled by each of the two detector dies.
Combining the two offset detector dies allows a sampling of the PSF with an effective detector pitch of \SI{\sim 1.2}{mm}.
The distance between the two dies will be less than \SI{1}{cm}.
Assuming the focal plane of the X-ray optics is placed directly between the two dies, the HPD of the point spread function will increase by only \SI{\sim 2}{\percent}.
The array diameter corresponds to an angular scale of about \ang{;8.8;}, significantly larger than the field of view of the X-ray optics, which eases the requirements on alignment stability as long as alignment knowledge can be maintained.

\subsubsection{Sensor readout}
TES arrays use multiplexing to minimize the thermal load and cryogenic complexity of wire connections to room temperature.
The development of the Microwave SQUID Multiplexer, which reads out array of microcalorimeters using microwave techniques~\cite{mates_etal_2011,mates_etal_2017}, increases the available measurement bandwidth from ${\sim}\SI{30}{MHz}$ (the intrinsic limit in previous multiplexing architectures) to the several \si{GHz} of bandwidth available on a single coaxial cable.

\begin{figure*}
    \centering
    \includegraphics[width=\textwidth]{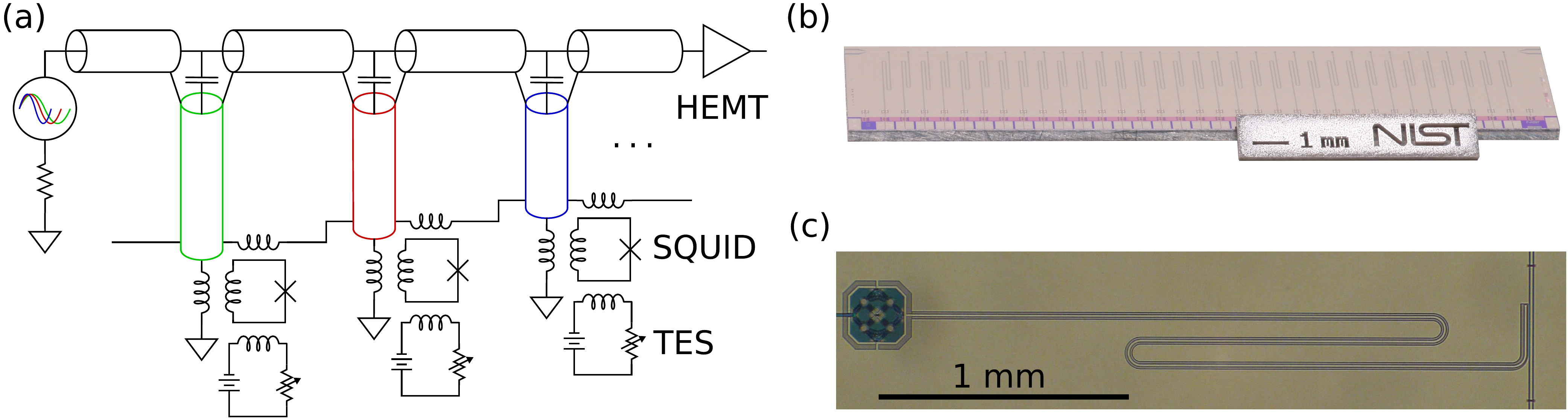}
  \caption{
    (a) Circuit schematic showing three channels of a microwave SQUID multiplexing circuit with TES microcalorimeters.
    (b) A photograph of a 33-channel microwave SQUID multiplexer chip used in the \sledgehammer instrument.
    (c) A close-up photograph showing quarter-wave microwave resonators capacitively coupled to a feedline. The resonators are terminated by inductively coupled rf-SQUIDs (left).
    The microwave SQUID multiplexer takes advantage of the large bandwidth provided by coaxial cables to significantly reduce the thermal load and design complexity of reading out large-format arrays of TES microcalorimeters, such as those used in \ascent.
    Reprinted from Mates et al.\ (2017)\cite{mates_etal_2017} with the permission of AIP Publishing.
    }
  \label{fig:squid-multiplexing}
\end{figure*}

In the Microwave SQUID Multiplexer~\cite{mates_2008}, each sensor is coupled to a high-Q, thin-film resonant circuit by an rf-SQUID that transduces current changes at the sensor to changes in inductive load on the resonator (Fig.~\ref{fig:squid-multiplexing}).
Multiple resonators, each with a unique frequency, are coupled to a single microwave feedline.
A sum of microwave tones (sine waves) is supplied to the feedline, each tone matched to the frequency of one resonator.
Changes in the current through a sensor will shift the center frequency of its resonator and thus change the amplitude and phase of the tone that propagates through the feedline.
All tones are amplified by a single shared cryogenic low-noise amplifier before returning to room temperature, where they are analyzed to extract the detector signals.
Signals from different sensors can easily be separated because they appear in modulation sidebands of their respective tones.
The first microcalorimeter array with microwave readout, \sledgehammer, demonstrated multiplexing factors of \num{128} with negligible resolution degradation, yielding a co-added resolution of \SI{55}{eV} at the \SI{97}{keV} gamma-ray peak (Figs.~\ref{fig:sledgehammer} and~\ref{fig:tes-image}).

\ascent will use the same 33-resonator microwave SQUID multiplexing chip designs used for \sledgehammer, shown in Fig.~\ref{fig:squid-multiplexing}(b).
Each resonator has a FWHM bandwidth of $\sim$\SI{300}{kHz}, and the resonances are spaced \SI{3}{MHz} apart.
Variations to the design place 32 resonators into each of eight \SI{125}{MHz} bands between \SI{5}{GHz} and \SI{6}{GHz}, yielding a total density of 256 detectors per GHz of available bandwidth.
To read out the 512 detectors used in \ascent, we will use 2 parallel pairs of coaxial cables, each reading out a separate set of resonators in the \SIrange{5}{6}{GHz} range.

The signals will be analyzed by four commercially available \textsl{ROACH2} Field Programmable Gate Array (FPGA) systems designed by the \textsl{CASPER} radio astronomy consortium~\cite{hickish_etal_2016}, each covering a bandwidth of \SI{512}{MHz} using commercially available DAC and ADC daughter boards.

\subsubsection{Calibration}
Achieving the best possible spectroscopic performance requires constant monitoring of the detector calibration, in order to be able to correct for calibration drift.
\ascent will carry a calibration source housed inside a tungsten enclosure outside a dedicated window in the cryostat, illuminating the detector array from behind.
The calibration of gamma-ray TESs drifts on timescales of minutes to hours.
Experience shows that the dominant contribution to short-term drift is correlated with the baseline and can be corrected for.
Therefore, it is necessary to obtain calibration spectra about once per hour to keep the systematic uncertainty due to residual drift below \SI{10}{eV}.
By collecting about 100 photons per calibration line, we can keep the statistical uncertainty of the calibration negligible. 
Two approaches to calibration are still under consideration: continuous illumination with a weak source or use of a strong source behind a shutter periodically illuminating the array for a brief period of time.

The advantage of continuous illumination is that no shutter mechanism is required and that no artificial dead-time is introduced.
This reduces mission complexity and, thus, risk.
However, care must be taken in the selection of calibration isotope or combination of isotopes.
Calibration lines should be close to the energy range of interest, but lines within the energy range of interest, including escape peaks, will cause unacceptable background.
Furthermore, emission lines above the energy range of interest will cause a background continuum due to Compton scattering in the detectors.
A preliminary Geant4~\cite{agostinelli_etal_2003,allison_etal_2006,allison_etal_2016} simulation study of calibration spectra obtained with a selection of viable sources revealed $^{155}$Eu with sufficiently strong lines at \SI{60}{keV} and \SI{86}{keV} as a candidate.

A sufficiently strong source behind a shutter will allow us to acquire calibration spectra at regular intervals over a short period of time.
Because calibration data are acquired at known time intervals during which science data collection will be suspended, it is possible to use calibration lines in the energy range of interest.
This has the advantage of mitigating the impact of non-linearities in the detector response.
The approach also reduces the continuum background due to high-energy lines from the calibration source, assuming sufficient shielding is possible.
The downside of this approach is that a mechanical shutter is required, adding complexity.
While details still need to be optimized, a preliminary estimate shows that a \SI{1}{min} calibration window every hour will be sufficient.
A possible calibration isotope is $^{227}$Ac which emits a large number of X-ray and gamma-ray lines in the energy range of interest.
This calibration method would result in an additional dead-time of~\SI{<2}{\percent}.

\subsubsection{Anti-coincidence detector}
In order to reduce the background, the cryostat section containing the focal plane instrumentation will almost entirely be enclosed in a ${\sim}$\SIrange{2}{3}{cm}-thick active bismuth-germanate (BGO) anti-coincidence shield.
Scintillation light due to particles interacting in the anti-coincidence shield will be detected by photomultiplier tubes (PMTs) or Silicon Photomultipliers (SiPMs).
Signals in these PMTs or SiPMs will produce a flag vetoing any triggers in the TES detector readout, significantly reducing the residual background, in addition to the passive shielding provided by the absorption of particles in the BGO.
The design will maximize the solid angle covered by the active shield.
The conservative solution is to place the active shield components at the outside of the cryostat, avoiding the difficulties associated with bringing the scintillator crystal and detectors from ambient temperatures and pressure to liquid helium temperatures and near vacuum.
In this case, passive tungsten shielding will be used inside the cryostat for the small solid angle portions not covered by the active shield outside the cryostat.
We are evaluating solutions with active shielding inside the cryostat which would result in a smaller shield and thus reduced cross sections for interactions with the background, and a reduced shield mass.
In both cases, the veto flag will be fed to the data acquisition and will be digitized along with the TES signals.

\subsection{X-ray optics}
\ascent achieves a large effective area in the \SIrange{65}{85}{keV} energy range using a dedicated multilayer-coated grazing incidence X-ray mirror.
The mirror consists of 213 nested shells in two reflection stages with a diameter of \SI{40}{cm}.
The innermost 110 shells will be coated with approximately 500 Ni/C layer pairs, while the remaining shells will be coated with roughly 200 Pt/C layer pairs.
Given the focal length of \SI{12}{m}, the design will limit incidence angles to $<\ang{0.23}$.
Reflectivity over a broad bandwidth at high X-ray energies is achieved by coating the shells with alternating layers of high-Z and low-Z material.
The \ascent optics are expected to achieve an angular resolution of \ang{;2;} half-power diameter (HPD).
The field of view of \ang{;5;} FWHM exceeds the angular size of \casA of \ang{;4;}.

\begin{figure}
  \centering
  \includegraphics[width=.7\linewidth]{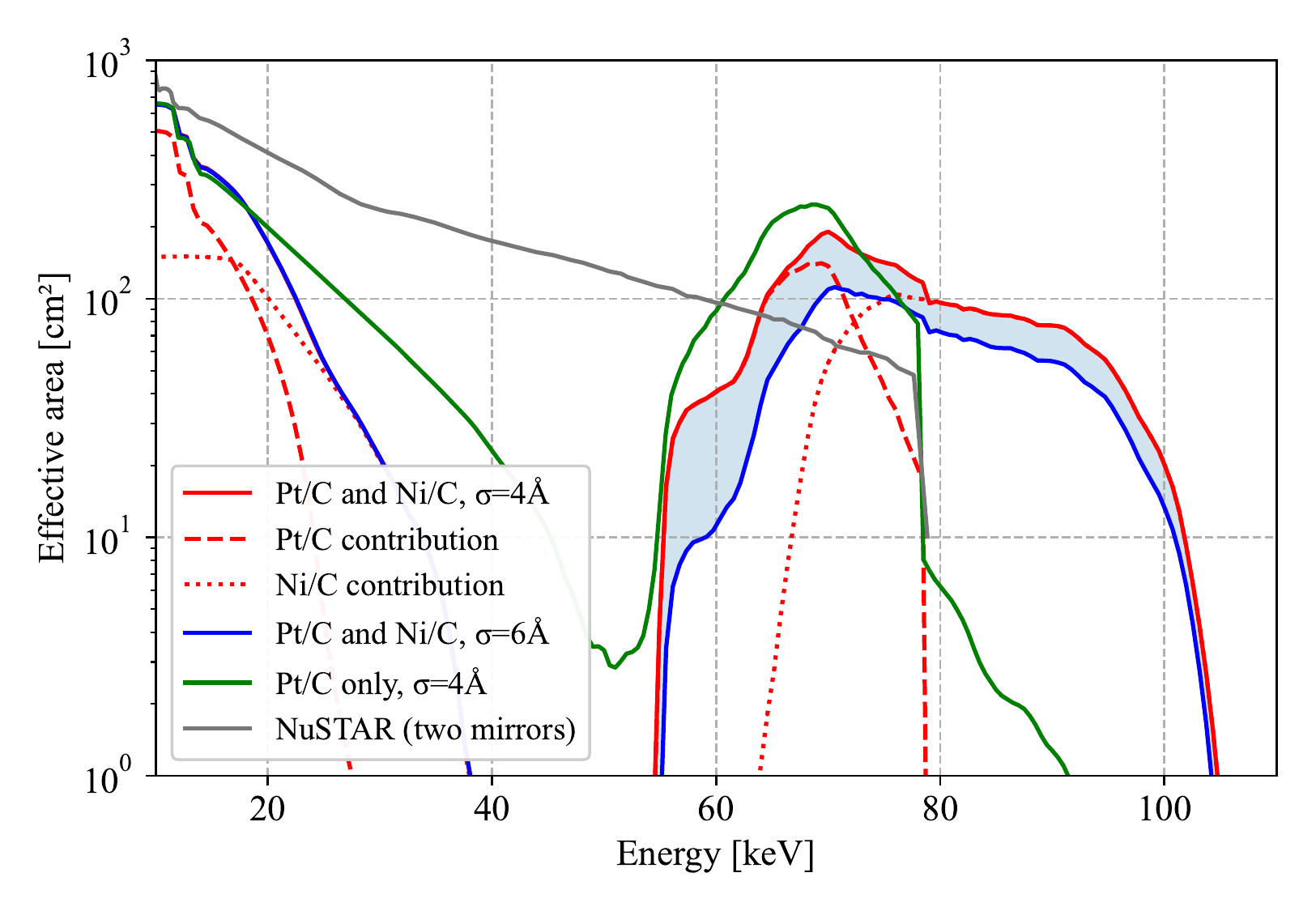}
  \caption{Comparison of the effective areas of mirror designs for \ascent and the two \nustar mirrors combined~\cite{harrison_etal_2013}.
  The \ascent design uses a combination of Ni/C multilayer coatings on the inner 110 shells and Pt/C on the outer shells.
  An alternative design using Pt/C layers on all shells is shown for comparison. The shaded region indicates the expected range of the surface roughness between $\sigma = \SI{6}{\angstrom}$ (worst case) and \SI{4}{\angstrom} (best case).
  The dashed and dotted red lines indicate the contribution of the Pt/C and Ni/C shells in the $\sigma = \SI{4}{\angstrom}$ case, respectively.
  }
  \label{fig:mirror_area}
\end{figure}

Most X-ray telescopes are designed to achieve a high collection area over a broad energy range.
The broadband design necessitates the deposition of a multilayer stack for soft X-rays on top of the stack for hard X-rays.
However, the thick soft X-ray layers absorb some of the higher-energy photons.
For \ascent, the multilayer design will be optimized for energies above \SI{60}{keV}, achieving there substantially higher reflectivities than a broadband X-ray multilayer coating.
Figure \ref{fig:mirror_area} shows the comparison of the collection areas of the \ascent and the two \nustar mirrors.
The platinum K absorption edge at \SI{78.395}{keV} limits the effective area at higher energies, preventing a purely Pt/C multilayer mirror from properly observing the \ti line at \SI{78.36}{keV}, especially when that line is blueshifted.
Therefore, in addition to an optimized multilayer structure, the \ascent optics will use Ni/C coatings on the innermost 110 out of 213 nested shells.
The Ni/C design requires about 500 layer pairs, 
while the Pt/C design requires about 200 layer pairs, which limits the number of foils that can be fabricated with this method within the project timeline.
A surface roughness of \SIrange{4}{6}{\angstrom} of the multilayer coatings is expected.
The \ascent mirror achieves collection areas of \SI{>100}{cm^2} in the \SIrange{65}{80}{keV} band.
Due to the optimization of the layer structure for \ti observations, the effective area between \SI{\sim 30}{keV} and \SI{55}{keV} is very small.
The low-energy reflectivity is not due to Bragg reflection on the multilayer, but due to total external reflection on the surface.
The Ni/C multilayer combination has been studied by several groups in the past (\eg Spiga et al., 2004\cite{spiga_etal_2004}) and is considered a top candidate material for future missions, such as \hexp, to extend their energy band out to~\SI{200}{keV}~\cite{madsen_etal_2018}.

\subsection{Cryostat}
\begin{figure}
  \centering
  \includegraphics[width=.6\linewidth]{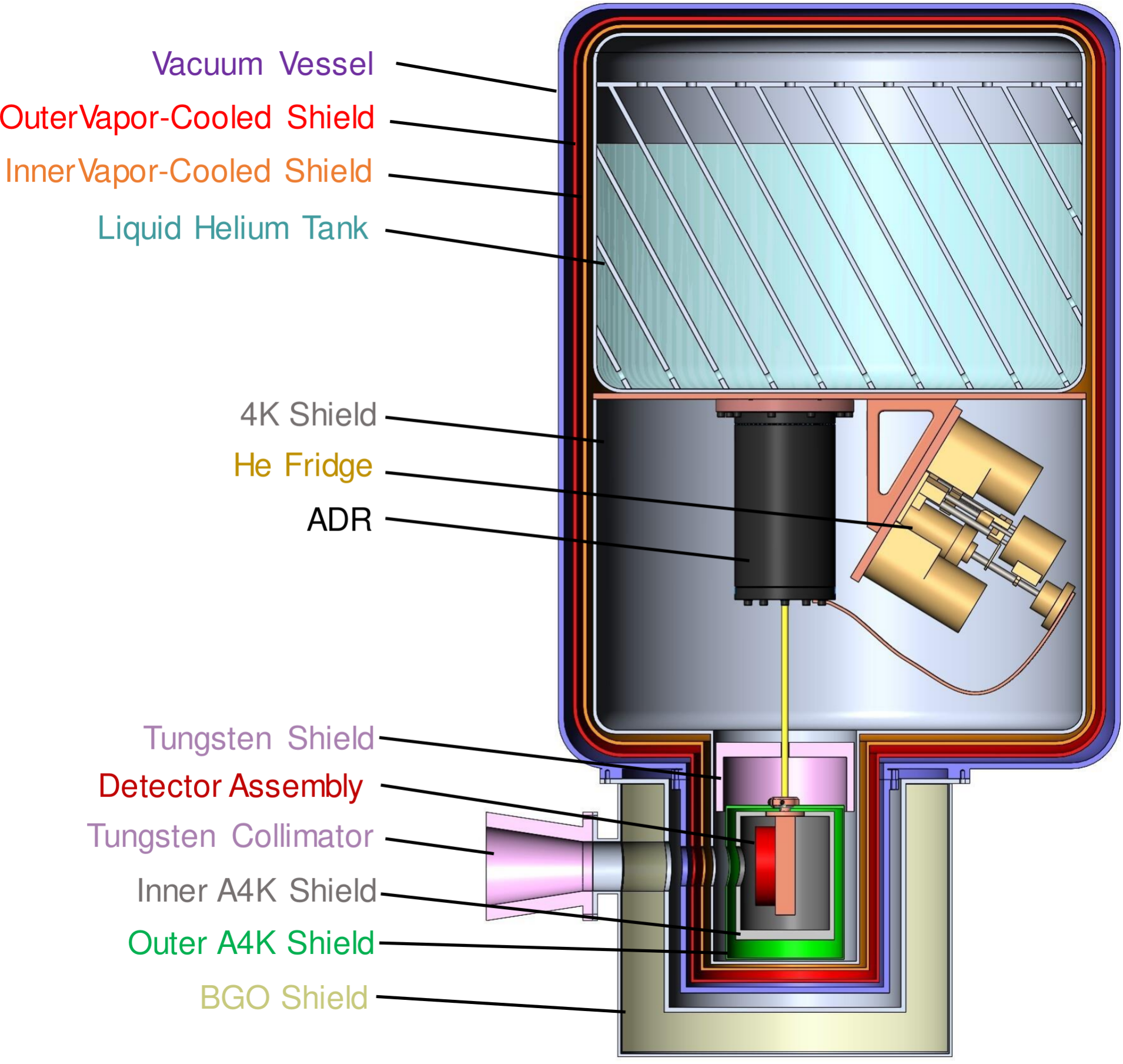}
  \caption{Conceptual design of the \ascent cryostat, which provides a \SI{70}{mK} base temperature for the detectors.  The detectors are housed in a protruding snout to minimize the mass of the shielding.}
  \label{fig:cryo}
\end{figure}
The \ascent cryogenic system cools the detector assembly to a nominal base temperature of \SI{70}{mK}.
Our baseline design foresees to use the cryostat architecture of Fig.~\ref{fig:cryo} that uses a commercial adiabatic demagnetization refrigerator (ADR) coupled to a closed-cycle \SI{300}{mK} refrigerator and a liquid helium bath.
The~\SI{65}{L} tank is designed to cool the detectors for up to 14 days, more than sufficient for the expect duration of the balloon flights from Sweden to Northern Canada.
An absolute pressure regulator maintains the tank near atmospheric pressure to provide a~\SI{4}{K} thermal bath, and internal baffles minimize sloshing to prevent resonances with the pointing system.
Counterweights on the pointing system maintain payload balance as the motion of the liquid during elevation changes and cryogen boil-off shift the center of mass. 

The main liquid helium tank is insulated by two vapor-cooled shields.
As cold gas boils away from the liquid helium reservoir, it flows through stainless steel pipes connected to the shields through low-impedance heat exchangers.
Since the helium boil off rate is proportional to the thermal load, negative feedback enforces temperature stability within the cryostat.
G-10 trusses mechanically support each stage while maintaining sufficient thermal insulation.
Liquid nitrogen cooling was considered as an alternative for the vapor-cooled shields, but would have led to a heavier and more complex design due to the additional cryogen tank.

A~\SI{300}{mK} temperature stage is provided by a multi-stage closed-cycle He-4/He-3 sorption refrigerator coupled to the cryostat's liquid helium tank.
This serves as the launching point for the ADR, allowing a lower magnetic field strength and thus lower power consumption than launching directly from~\SI{4}{K}.  The~\SI{300}{mK} stage also intercepts the parasitic load from the wiring and mechanical support structures to reduce the cooling power requirement on the lowest temperature stage.
Similar refrigerators have been successfully used by many different balloon-borne cryogenic systems (\eg~SPIDER\cite{gudmundsson_etal_2015}, EBEX\cite{EBEX_cryo}, BLAST-Pol\cite{blast_cryo}, and BOOMERANG\cite{boomerang_crill}).

The detector assembly is maintained at a base temperature of \SI{70}{mK} by a commercially available ADR using a single ferric ammonium alum (FAA) salt pill~\cite{ADR_FAA_ref} launched from the~\SI{300}{mK} stage to provide~\SI{1}{\micro\watt} of cooling power with~\SI{120}{mJ} cooling capacity.
A~3-hour regeneration cycle will be performed once every 24 hours, providing roughly~\SI{90}{\percent} observing efficiency.
For optimal operation, the TES detectors and SQUID multiplexer chips must be protected from external magnetic fields.
The entire detector package assembly containing both detector dies and all SQUID multiplexer chips will therefore be enclosed in a two-layer magnetic shield, incorporated within the cryostat and maintained at cryogenic temperatures.
These reduce the magnetic flux in the SQUIDs due to Earth's magnetic field by about two orders of magnitude.

As an alternative cooling option, we are currently evaluating the performance of a mini Dilution Refrigerator.\cite{chase_dilutor}
First measurements in the lab indicate that the mini Dilution Refrigerator is well suited for this application, offering continuous cooling to \SI{80}{mK} temperatures.
The cooling power does not change significantly for elevation changes of $\pm\ang{30}$, enabling its use for \ascent.

\subsection{Optical bench and gondola}
X-ray optics and cryostat will be supported by a \SI{12}{m}-long optical bench pointed by NASA's Wallops Arc-Second Pointer (WASP) system.
The optical bench consists of three sections made of carbon fiber tubes glued to Aluminum joints.
The design results in an extremely stiff truss and is similar to previous balloon-borne telescopes \xcalibur and \xlcalibur~\cite{kislat_etal_2017, abarr_etal_2020b}.
For example, the \SI{8}{m} long truss of \xcalibur achieved a stability of \SI{<1.5}{mm} of the focal point during most of the flight~\cite{abarr_etal_2021}.
The X-ray mirror, star tracker, and fiber-optic gyro of the WASP will be mounted to an Al honeycomb panel at the front end of the truss, and the focal plane instrumentation will be attached to an Al honeycomb panel at the rear end of the truss. 

The stiffness of the optical bench fulfills two requirements.
First, the WASP pointing system requires that the lowest-frequency vibration mode of the pointed body exceeds~\SI{10}{Hz}.
Second, we require a motion of the focal spot \SI{<3}{mm} in order to ensure the entire image is always contained in the detector array.
A focal spot motion of \SI{3}{mm} corresponds to \ang{;;50} pointing error.
In order to reduce the resulting degradation of the point spread function, \ascent will use an alignment monitoring system similar to \xcalibur\cite{abarr_etal_2021}.
The system uses an optical camera mounted in the central bore of the X-ray optics observing a pattern of LEDs mounted to the entrance window of the detector.
On a \SI{12}{m} truss, it measures the alignment with a precision of \SI{0.15}{mm} or \ang{;;2.5}, negligible compared to the point spread function of the optics.

An aluminum gondola suspended from the balloon supports the truss pointed by the WASP.
The WASP points the truss in pitch and within a limited yaw range with respect to the gondola.
Coarse pointing in yaw is achieved using a standard NASA balloon rotator coupling the gondola to the balloon.
Absolute pointing information is provided by a star tracker system specially developed for balloon flight applications.
This system achieves a pointing stability of~${<}\ang{;;1}$ and an absolute pointing accuracy of~${\sim}\ang{;;15}$.

\section{Expected performance}\label{sec:sensitivity}
During a balloon flight from Kiruna, Sweden, \ascent will observe \casA for approximately \SI{500}{ksec} at an elevation of about \num{36}--\ang{82}. 
We envision multiple northern hemisphere flights in order to attain longer total observation times.

In order to estimate \ascent's sensitivity, we simulated the detector in  Geant4 as two stacked arrays of Sn absorbers with the layout as shown in Fig.~\ref{fig:tes-image}c and a thickness of \SI{0.59}{mm}.
Input spectra were folded with the mirror effective area based on the Ni/C multilayer mirror shown in Fig.~\ref{fig:mirror_area} and energy-dependent atmospheric absorption at a balloon altitude of \SI{125000}{ft} corresponding to an overburden of \SI{2.9}{\g\per\square\cm}.
Photons were distributed according to the mirror point spread function (PSF) of ASTRO-H HXT\cite{matsumoto_etal_2016}, which is similar to the expected \ascent PSF.

We estimate the background using measurements made during the Antarctic flight of \xcalibur~\cite{abarr_etal_2021} and taking into account improvements to the anticoincidence shield, which will reduce this background by a factor of \num{\sim 10}~\cite{abarr_etal_2020b}.
To account for the difference in detector size we scale with area and square root of thickness.
This results in a background rate of \SI{\sim 1.7e-6}{\per\s\per\keV} per TES detector at \SI{68}{keV}.
In the future, this estimate can be refined using data from an upcoming test flight of a small TES array scheduled for the fall of 2023, as well as with the help of detailed Monte Carlo simulations.\cite{shirazi_etal_2022}

\begin{figure}
  \centering%
  \includegraphics[width=.6\textwidth]{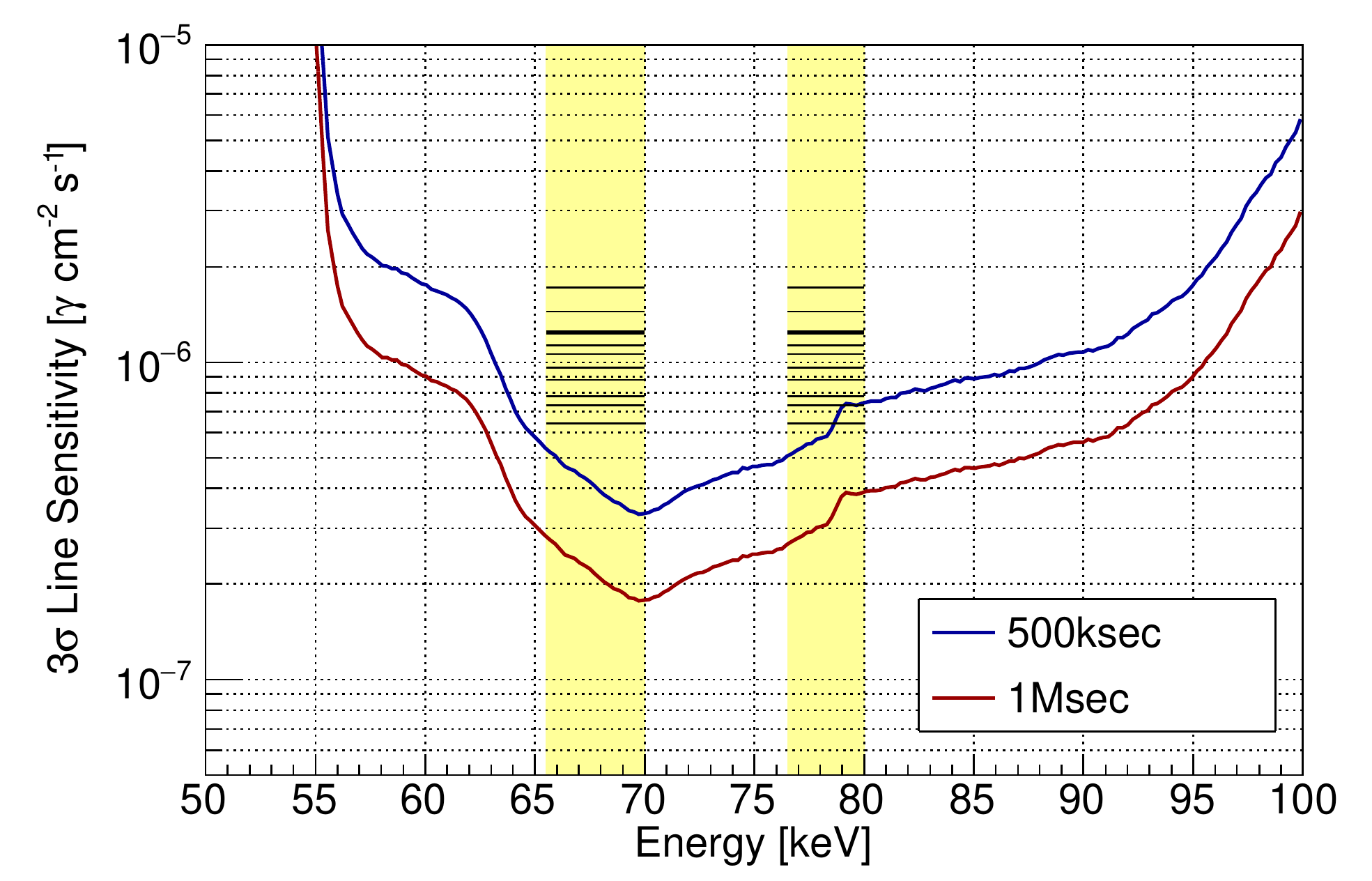}%
  \caption{%
    \ascent narrow line sensitivity as a function of energy, demonstrating \ascent's ability to detect \ti in various regions of \casA with a single \SI{500}{ksec} flight.
    The vertical yellow bands indicate energy ranges of interest and the horizontal black bars are the flux levels measured by \nustar.\cite{grefenstette_etal_2017}
  }
  \label{fig:narrow-line-sensitivity}
\end{figure}

In the sensitivity calculation, we weighted events in each detector by the expected signal-to-background ratio for a point-like source according to Ref.~\cite{barlow_1987}.
Figure~\ref{fig:narrow-line-sensitivity} shows the expected narrow line sensitivity of \ascent.
NuSTAR detected lines with fluxes ranging from \SI{6e-7}{\per\square\centi\meter\per\second} to \SI{1.7e-6}{\per\square\centi\meter\per\second}.
\ascent's energy resolution of \SI{67}{eV} FWHM will allow us to determine the velocity of \ti ejecta with a FWHM accuracy of \SI{270}{\km\per\s}, compared to \nustar's FWHM accuracy of \SI[group-separator={,}]{3600}{\km\per\second}.

\begin{figure}
  \centering%
  \includegraphics[width=.6\textwidth]{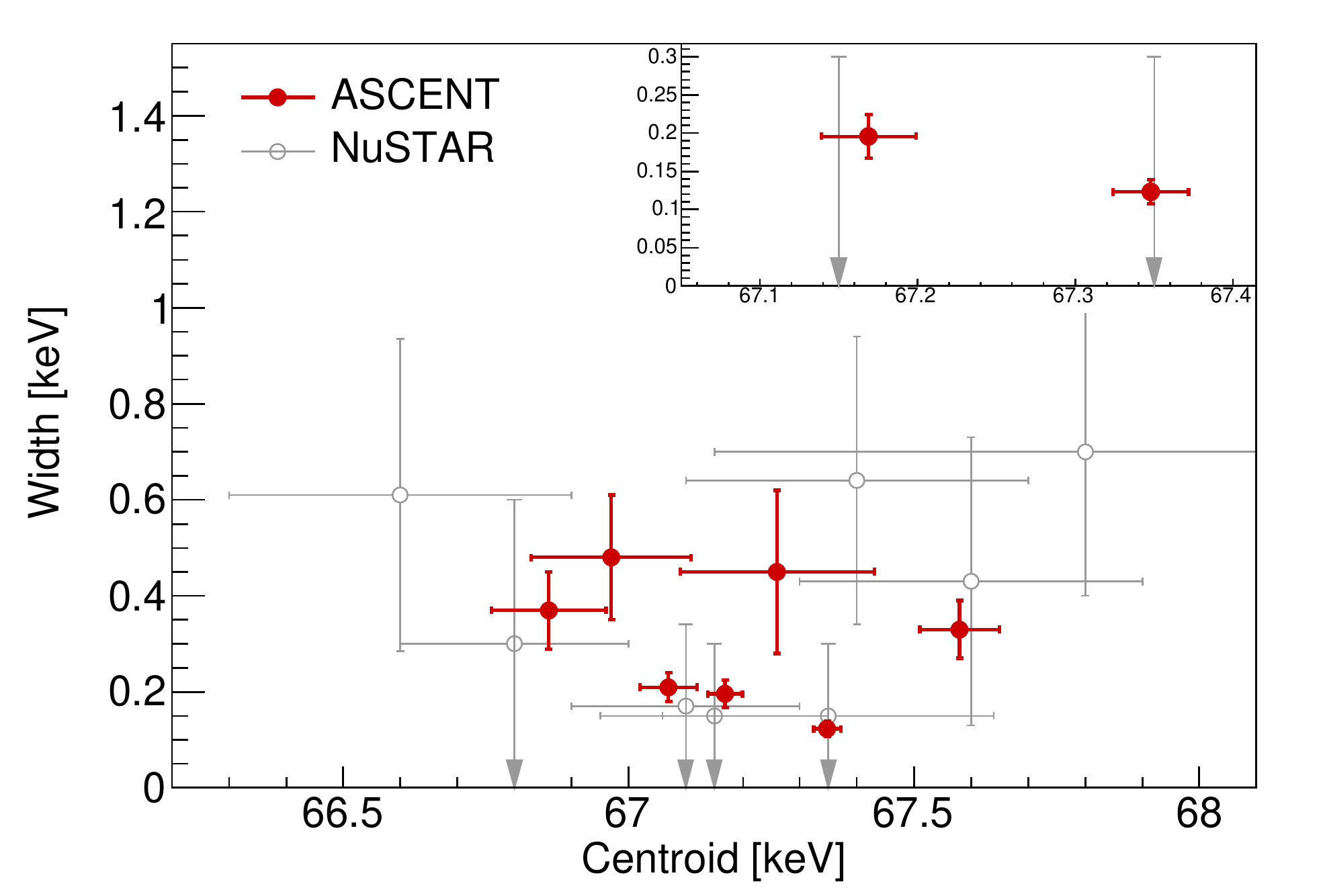}%
  \caption{%
    Expected \ascent results for the measurements of \ti emission lines in the 8 \ti-bright regions analyzed in Ref.~\cite{grefenstette_etal_2017} compared to the \nustar results. Simulations assume the Gaussian line parameters measured by \nustar. In regions where \nustar only set an upper limit on the line width, a smaller value is assumed. The inset shows two lines with assumed widths of \SI{150}{eV} and \SI{100}{eV}, respectively. The \ascent centroid and line width uncertainties are a factor \numrange{2}{10} smaller than \nustar's.}
  \label{fig:linemeasurements}
\end{figure}

From each flight we expect highly significant detections of the \ti emission from \casA of $11\,\sigma$ and $6.5\,\sigma$ of the \SI{67.9}{keV} and \SI{78.4}{keV} lines, respectively, when summing over all bright spots.
Particularly interesting will be the energy spectrum from \textit{Region~1} in Fig.~\ref{fig:casa}, which we expect to detect with $5\,\sigma$ at \SI{67.9}{keV} and with $3\,\sigma$ at \SI{78.4}{keV}.
The advantage of \ascent is greatest in regions where the width of the \ti lines is small compared to \nustar's energy resolution.
A single \SI{500}{ksec} observation with \ascent will improve measurements in all regions where \nustar detected \ti, except two where the lines are broadened and very weak.
We estimate that \ascent will improve the line centroid and width measurements compared to previous results by a factor of \numrange{2}{20} and \numrange{2}{10}, respectively (Fig.~\ref{fig:linemeasurements}).
These measurements will significantly improve the 3D localization of the \ti ejecta and result in much tighter constraints on the local \nickel/\ti ratio.
These improvements in constraining the local ratio will greatly increase our knowledge of the nuclear production pathways in the supernova explosion.

\section{Summary and Outlook}\label{sec:summary}
Core collapse supernovae are considered to be a significant source of mid-Z elements in the Galaxy.
Despite significant theoretical and observational progress in the last few decades, many details of the explosion are still poorly understood.
One observational approach to gaining new insights is to study the distribution of elements in supernova remnants.
The isotope \ti is of particular interest because it is produced in the innermost regions of the supernova engine, and it is radioactive with a half-life long enough to be observable in historic supernova remnants.

In this paper we present a concept for a balloon-borne hard X-ray telescope called \ascent designed to study the nuclear emission lines from \casA and SN~1987A, as well as potentially other remnants that may be discovered by \cosi.
\ascent uses an array of transition edge sensor (TES) microcalorimeter detectors as its focal plane instrument, improving spectral resolution by more than an order of magnitude over existing semiconductor detectors at gamma-ray energies.
Observations with \ascent will significantly improve 3D maps of \ti in \casA, and can deliver detailed spectra of \ti from SN~1987A.

\ascent will also demonstrate the viability of hard X-ray TES technology for a future space mission.
The energy resolution of an \ascent-type mission would benefit all spectral studies of a \nustar follow-up, \eg broadband observations of Active Galactic Nuclei (AGN) covering the lines of the soft excess emission, the Fe K$\alpha$ emission, and the Compton hump emission.
The broadband results obtained for a sample of ${\sim}100$ of AGNs can be used to calibrate the spin measurements of ESA's \textsl{ATHENA} mission~\cite{smith_etal_2016,barcons_etal_2017} and the proposed NASA \textsl{Lynx} mission~\cite{gaskin_etal_2015,bandler_etal_2016, lynx}.
The \ascent detector technology could be used for example as the focal plane detector of the  \hexp mission.\cite{madsen_etal_2018}
While \ascent has not been selected when first proposed, the team continues to pursue the project and is planning to repropose the mission after maturing detector, cryostat, and X-ray optics technologies.

\section*{Acknowledgments}
FK is grateful for support by the Faculty Development Grant program of UNH.
The work by CLF is supported by the US Department of Energy through the Los Alamos National Laboratory. Los Alamos National Laboratory is operated by Triad National Security, LLC, for the National Nuclear Security Administration of the U.S.\ Department of Energy (Contract No.\ 89233218CNA000001).
HK and JN acknowledge NASA support under grant 80NSSC21K1817.
HK acknowledges NASA support under grants 80NSSC18K0264, 80NSSC22K1291, and NNX16AC42G.


\bibliography{ascent}   

\begin{thebibliography}{10}

\bibitem{burbidge_etal_1957}
E.~M. {Burbidge}, G.~R. {Burbidge}, W.~A. {Fowler}, {\em et~al.}, ``{Synthesis
  of the Elements in Stars},'' {\em Rev. Mod. Phys.} {\bf 29}, 547--650
  (1957).

\bibitem{matz_etal_1988}
S.~M. {Matz}, G.~H. {Share}, M.~D. {Leising}, {\em et~al.}, ``{Gamma-ray line
  emission from SN1987A},'' {\em Nature} {\bf 331}, 416--418  (1988).

\bibitem{choi_etal_1998}
B.-G. {Choi}, G.~R. {Huss}, G.~J. {Wasserburg}, {\em et~al.}, ``{Presolar
  Corundum and Spinel in Ordinary Chondrites: Origins from AGB Stars and a
  Supernova},'' {\em Science} {\bf 282}, 1284  (1998).

\bibitem{schulte_etal_2021}
J.~{Schulte}, M.~{Bose}, P.~A. {Young}, {\em et~al.}, ``{Three-dimensional
  Supernova Models Provide New Insights into the Origins of Stardust},'' {\em
  Astrophys. J.} {\bf 908}, 38  (2021).

\bibitem{amari_etal_1992}
S.~{Amari}, P.~{Hoppe}, E.~{Zinner}, {\em et~al.}, ``{Interstellar SiC with
  Unusual Isotopic Compositions: Grains from a Supernova?},'' {\em Astrophys.
  J. Lett.} {\bf 394}, L43  (1992).

\bibitem{herant_benz_1992}
M.~{Herant} and W.~{Benz}, ``{Postexplosion Hydrodynamics of SN 1987A},'' {\em
  Astrophys. J.} {\bf 387}, 294  (1992).

\bibitem{ahmad_etal_2006}
I.~{Ahmad}, J.~P. {Greene}, E.~F. {Moore}, {\em et~al.}, ``{Improved
  measurement of the Ti44 half-life from a 14-year long study},'' {\em Phys.
  Rev. C} {\bf 74}, 065803  (2006).

\bibitem{chen_etal_2011}
J.~{Chen}, B.~{Singh}, and J.~A. {Cameron}, ``{Nuclear Data Sheets for A =
  44},'' {\em Nucl. Data Sheets} {\bf 112}, 2357--2495  (2011).

\bibitem{clayton_etal_1969}
D.~D. {Clayton}, S.~A. {Colgate}, and G.~J. {Fishman}, ``{Gamma-Ray Lines from
  Young Supernova Remnants},'' {\em Astrophys. J.} {\bf 155}, 75  (1969).

\bibitem{ullom_bennett_2015}
J.~N. {Ullom} and D.~A. {Bennett}, ``{Review of superconducting transition-edge
  sensors for x-ray and gamma-ray spectroscopy},'' {\em Supercond. Sci. Tech.}
  {\bf 28}, 084003  (2015).

\bibitem{mates_etal_2017}
J.~A.~B. {Mates}, D.~T. {Becker}, D.~A. {Bennett}, {\em et~al.},
  ``{Simultaneous readout of 128 X-ray and gamma-ray transition-edge
  microcalorimeters using microwave SQUID multiplexing},'' {\em Appl. Phys.
  Lett.} {\bf 111}, 062601  (2017).

\bibitem{harrison_etal_2013}
F.~A. {Harrison}, W.~W. {Craig}, F.~E. {Christensen}, {\em et~al.}, ``{The
  Nuclear Spectroscopic Telescope Array (NuSTAR) High-energy X-Ray Mission},''
  {\em Astrophys. J.} {\bf 770}, 103  (2013).

\bibitem{boggs_etal_2015}
S.~E. {Boggs}, F.~A. {Harrison}, H.~{Miyasaka}, {\em et~al.}, ``{$^{44}$Ti
  gamma-ray emission lines from SN1987A reveal an asymmetric explosion},'' {\em
  Science} {\bf 348}, 670--671  (2015).

\bibitem{grebenev_etal_2012}
S.~A. {Grebenev}, A.~A. {Lutovinov}, S.~S. {Tsygankov}, {\em et~al.},
  ``{Hard-X-ray emission lines from the decay of $^{44}$Ti in the remnant of
  supernova 1987A},'' {\em Nature} {\bf 490}, 373--375  (2012).

\bibitem{iyudin_etal_1994}
A.~F. {Iyudin}, R.~{Diehl}, H.~{Bloemen}, {\em et~al.}, ``{COMPTEL observations
  of Ti-44 gamma-ray line emission from CAS A},'' {\em Astron. Astrophys.} {\bf
  284}, L1--L4  (1994).

\bibitem{vink_etal_2001}
J.~{Vink}, J.~M. {Laming}, J.~S. {Kaastra}, {\em et~al.}, ``{Detection of the
  67.9 and 78.4 keV Lines Associated with the Radioactive Decay of $^{44}$Ti in
  Cassiopeia A},'' {\em Astrophys. J. Lett.} {\bf 560}, L79--L82  (2001).

\bibitem{grefenstette_etal_2014}
B.~W. {Grefenstette}, F.~A. {Harrison}, S.~E. {Boggs}, {\em et~al.},
  ``{Asymmetries in core-collapse supernovae from maps of radioactive $^{44}$Ti
  in Cassiopeia A},'' {\em Nature} {\bf 506}, 339--342  (2014).

\bibitem{iyudin_etal_1998}
A.~F. {Iyudin}, V.~{Sch{\"o}nfelder}, K.~{Bennett}, {\em et~al.}, ``{Emission
  from $^{44}$Ti associated with a previously unknown Galactic supernova},''
  {\em Nature} {\bf 396}, 142--144  (1998).

\bibitem{troja_etal_2014}
E.~{Troja}, A.~{Segreto}, V.~{La Parola}, {\em et~al.}, ``{Swift/BAT Detection
  of Hard X-Rays from Tycho's Supernova Remnant: Evidence for Titanium-44},''
  {\em Astrophys. J. Lett.} {\bf 797}, L6  (2014).

\bibitem{tsygankov_etal_2016}
S.~S. {Tsygankov}, R.~A. {Krivonos}, A.~A. {Lutovinov}, {\em et~al.},
  ``{Galactic survey of $^{44}$Ti sources with the IBIS telescope onboard
  INTEGRAL},'' {\em Mon. Not. Royal Astron. Soc.} {\bf 458}, 3411--3419
  (2016).

\bibitem{lopez_etal_2015}
L.~A. {Lopez}, B.~W. {Grefenstette}, S.~P. {Reynolds}, {\em et~al.}, ``{A
  Spatially Resolved Study of the Synchrotron Emission and Titanium in Tycho's
  Supernova Remnant Using NuSTAR},'' {\em Astrophys. J.} {\bf 814}, 132
  (2015).

\bibitem{tomsick_etal_2021}
J.~A. {Tomsick}, S.~E. {Boggs}, A.~{Zoglauer}, {\em et~al.}, ``{The Compton
  Spectrometer and Imager Project for MeV Astronomy},'' in {\em Proc. 37th
  International Cosmic Ray Conference},   (2021).

\bibitem{woosley_weaver_1994}
S.~E. {Woosley} and T.~A. {Weaver}, ``{Sub--Chandrasekhar Mass Models for Type
  IA Supernovae},'' {\em Astrophys. J.} {\bf 423}, 371  (1994).

\bibitem{crocker_etal_2017}
R.~M. {Crocker}, A.~J. {Ruiter}, I.~R. {Seitenzahl}, {\em et~al.}, ``{Diffuse
  Galactic antimatter from faint thermonuclear supernovae in old stellar
  populations},'' {\em Nature Astronomy} {\bf 1}, 0135  (2017).

\bibitem{renaud_etal_2006}
M.~{Renaud}, J.~{Vink}, A.~{Decourchelle}, {\em et~al.}, ``{The Signature of
  $^{44}$Ti in Cassiopeia A Revealed by IBIS/ISGRI on INTEGRAL},'' {\em
  Astrophys. J. Lett.} {\bf 647}, L41--L44  (2006).

\bibitem{siegert_etal_2015}
T.~{Siegert}, R.~{Diehl}, M.~G.~H. {Krause}, {\em et~al.}, ``{Revisiting
  INTEGRAL/SPI observations of $^{44}$Ti from Cassiopeia A},'' {\em Astron.
  Astrophys.} {\bf 579}, A124  (2015).

\bibitem{young_etal_2006}
P.~A. {Young}, C.~L. {Fryer}, A.~{Hungerford}, {\em et~al.}, ``{Constraints on
  the Progenitor of Cassiopeia A},'' {\em Astrophys. J.} {\bf 640}, 891--900
  (2006).

\bibitem{grefenstette_etal_2017}
B.~W. {Grefenstette}, C.~L. {Fryer}, F.~A. {Harrison}, {\em et~al.}, ``{The
  Distribution of Radioactive $^{44}$Ti in Cassiopeia A},'' {\em Astrophys. J.}
  {\bf 834}, 19  (2017).

\bibitem{magkotsios_etal_2010}
G.~{Magkotsios}, F.~X. {Timmes}, A.~L. {Hungerford}, {\em et~al.}, ``{Trends in
  $^{44}$Ti and $^{56}$Ni from Core-collapse Supernovae},'' {\em Astrophys. J.
  Suppl. Ser.} {\bf 191}, 66--95  (2010).

\bibitem{koo_etal_2018}
B.-C. {Koo}, H.-J. {Kim}, Y.-H. {Lee}, {\em et~al.}, ``{A Deep Near-infrared
  [Fe II]+[Si I] Emission Line Image of the Supernova Remnant Cassiopeia A},''
  {\em Astrophys. J.} {\bf 866}, 139  (2018).

\bibitem{hammer_etal_2010}
N.~J. {Hammer}, H.~T. {Janka}, and E.~{M{\"u}ller}, ``{Three-dimensional
  Simulations of Mixing Instabilities in Supernova Explosions},'' {\em
  Astrophys. J.} {\bf 714}, 1371--1385  (2010).

\bibitem{ono_etal_2013}
M.~{Ono}, S.~{Nagataki}, H.~{Ito}, {\em et~al.}, ``{Matter Mixing in Aspherical
  Core-collapse Supernovae: A Search for Possible Conditions for Conveying
  $^{56}$Ni into High Velocity Regions},'' {\em Astrophys. J.} {\bf 773}, 161
  (2013).

\bibitem{vance_etal_2020}
G.~S. {Vance}, P.~A. {Young}, C.~L. {Fryer}, {\em et~al.}, ``{Titanium and Iron
  in the Cassiopeia A Supernova Remnant},'' {\em Astrophys. J.} {\bf 895}, 82
  (2020).

\bibitem{willingale_etal_2002}
R.~{Willingale}, J.~A.~M. {Bleeker}, K.~J. {van der Heyden}, {\em et~al.},
  ``{X-ray spectral imaging and Doppler mapping of Cassiopeia A},'' {\em
  Astron. Astrophys.} {\bf 381}, 1039--1048  (2002).

\bibitem{delaney_etal_2010}
T.~{DeLaney}, L.~{Rudnick}, M.~D. {Stage}, {\em et~al.}, ``{The
  Three-dimensional Structure of Cassiopeia A},'' {\em Astrophys. J.} {\bf
  725}, 2038--2058  (2010).

\bibitem{2006ApJS..163..335F}
C.~L. {Fryer} and A.~{Kusenko}, ``{Effects of Neutrino-driven Kicks on the
  Supernova Explosion Mechanism},'' {\em Astrophys. J. Suppl. Ser.} {\bf 163},
  335--343  (2006).

\bibitem{1994ApJ...435..339H}
M.~{Herant}, W.~{Benz}, W.~R. {Hix}, {\em et~al.}, ``{Inside the Supernova: A
  Powerful Convective Engine},'' {\em Astrophys. J.} {\bf 435}, 339  (1994).

\bibitem{1995PhR...256..117H}
M.~{Herant}, ``{The convective engine paradigm for the supernova explosion
  mechanism and its consequences.},'' {\em Phys. Rep.} {\bf 256}, 117--133
  (1995).

\bibitem{socrates_etal_2005}
A.~{Socrates}, O.~{Blaes}, A.~{Hungerford}, {\em et~al.}, ``{The Neutrino
  Bubble Instability: A Mechanism for Generating Pulsar Kicks},'' {\em
  Astrophys. J.} {\bf 632}, 531--562  (2005).

\bibitem{ellinger_etal_2012}
C.~I. {Ellinger}, P.~A. {Young}, C.~L. {Fryer}, {\em et~al.}, ``{A Case Study
  of Small-scale Structure Formation in Three-dimensional Supernova
  Simulations},'' {\em Astrophys. J.} {\bf 755}, 160  (2012).

\bibitem{stuchlik_2015}
D.~W. {Stuchlik}, ``{The Wallops Arc Second Pointer -- A Balloon Borne Fine
  Pointing System},'' in {\em AIAA Balloon Systems Conference},  {\em Proc.
  AIAA AVIATION Forum}  (2015).

\bibitem{aharonian_etal_2016}
F.~{Aharonian}, H.~{Akamatsu}, F.~{Akimoto}, {\em et~al.}, ``{The quiescent
  intracluster medium in the core of the Perseus cluster},'' {\em Nature} {\bf
  535}, 117--121  (2016).

\bibitem{smith_etal_2016}
S.~J. {Smith}, J.~S. {Adams}, S.~R. {Bandler}, {\em et~al.}, ``{Transition-edge
  sensor pixel parameter design of the microcalorimeter array for the x-ray
  integral field unit on Athena},'' in {\em Space Telescopes and
  Instrumentation 2016: Ultraviolet to Gamma Ray},  {\em Proc. SPIE} {\bf
  9905}, 99052H  (2016).

\bibitem{barcons_etal_2017}
X.~{Barcons}, D.~{Barret}, A.~{Decourchelle}, {\em et~al.}, ``{Athena: ESA's
  X-ray observatory for the late 2020s},'' {\em Astronomische Nachrichten} {\bf
  338}, 153--158  (2017).

\bibitem{bennett_etal_2012}
D.~A. {Bennett}, R.~D. {Horansky}, D.~R. {Schmidt}, {\em et~al.}, ``{A high
  resolution gamma-ray spectrometer based on superconducting
  microcalorimeters},'' {\em Rev. Sci. Instrum.} {\bf 83}, 093113--093113--14
  (2012).

\bibitem{winkler_etal_2015}
R.~{Winkler}, A.~S. {Hoover}, M.~W. {Rabin}, {\em et~al.}, ``{256-pixel
  microcalorimeter array for high-resolution {$\gamma$}-ray spectroscopy of
  mixed-actinide materials},'' {\em Nucl. Instrum. Meth. Phys. Res. A} {\bf
  770}, 203--210  (2015).

\bibitem{hoover_etal_2015}
A.~S. Hoover, E.~M. Bond, M.~P. Croce, {\em et~al.}, ``{Measurement of the
  240Pu/239Pu Mass Ratio Using a Transition-Edge-Sensor Microcalorimeter for
  Total Decay Energy Spectroscopy},'' {\em Anal. Chem.} {\bf 87}(7), 3996--4000
   (2015).

\bibitem{uhlig_etal_2013}
J.~{Uhlig}, W.~{Fullagar}, J.~N. {Ullom}, {\em et~al.}, ``{Table-Top Ultrafast
  X-Ray Microcalorimeter Spectrometry for Molecular Structure},'' {\em Phys.
  Rev. Lett.} {\bf 110}, 138302  (2013).

\bibitem{miaja_avila_etal_2016}
L.~{Miaja-Avila}, G.~C. {O'Neil}, Y.~I. {Joe}, {\em et~al.}, ``{Ultrafast
  Time-Resolved Hard X-Ray Emission Spectroscopy on a Tabletop},'' {\em Phys.
  Rev. X} {\bf 6}, 031047  (2016).

\bibitem{palosaari_etal_2016}
M.~R.~J. {Palosaari}, M.~{K{\"a}yhk{\"o}}, K.~M. {Kinnunen}, {\em et~al.},
  ``{Broadband Ultrahigh-Resolution Spectroscopy of Particle-Induced X Rays:
  Extending the Limits of Nondestructive Analysis},'' {\em Phys. Rev. Appl.}
  {\bf 6}, 024002  (2016).

\bibitem{okada_etal_2016}
S.~{Okada}, D.~A. {Bennett}, C.~{Curceanu}, {\em et~al.}, ``{First application
  of superconducting transition-edge sensor microcalorimeters to hadronic atom
  X-ray spectroscopy},'' {\em Prog. Theor. Exp. Phys.} {\bf 2016}, 091D01
  (2016).

\bibitem{oneil_etal_2017}
G.~C. O’Neil, L.~Miaja-Avila, Y.~I. Joe, {\em et~al.}, ``{Ultrafast
  Time-Resolved X-ray Absorption Spectroscopy of Ferrioxalate Photolysis with a
  Laser Plasma X-ray Source and Microcalorimeter Array},'' {\em J. Phys. Chem.
  Lett.} {\bf 8}(5), 1099--1104  (2017).

\bibitem{doriese_etal_2017}
W.~B. {Doriese}, P.~{Abbamonte}, B.~K. {Alpert}, {\em et~al.}, ``{A practical
  superconducting-microcalorimeter X-ray spectrometer for beamline and
  laboratory science},'' {\em Rev. Sci. Instrum.} {\bf 88}, 053108  (2017).

\bibitem{titus_etal_2017}
C.~J. {Titus}, M.~L. {Baker}, S.~J. {Lee}, {\em et~al.}, ``{L-edge spectroscopy
  of dilute, radiation-sensitive systems using a transition-edge-sensor
  array},'' {\em J. Chem. Phys.} {\bf 147}, 214201  (2017).

\bibitem{irwin_hilton_2005}
K.~D. {Irwin} and G.~C. {Hilton}, {\em {Transition-Edge Sensors}}, 63  (2005).

\bibitem{bacrania_etal_2009}
M.~K. Bacrania, A.~S. Hoover, P.~J. Karpius, {\em et~al.}, ``{Large-Area
  Microcalorimeter Detectors for Ultra-High-Resolution X-Ray and Gamma-Ray
  Spectroscopy},'' {\em IEEE Trans. Nucl. Sci.} {\bf 56}, 2299--2302  (2009).

\bibitem{weber_2020}
J.~Weber, K.~M. Morgan, D.~Yan, {\em et~al.}, ``Development of a
  transition-edge sensor bilayer process providing new modalities for critical
  temperature control,'' {\em Superconductor Science and Technology}   (2020).

\bibitem{mates_etal_2011}
J.~A.~B. {Mates}, K.~D. {Irwin}, L.~R. {Vale}, {\em et~al.}, ``{Flux-Ramp
  Modulation for SQUID Multiplexing},'' {\em J. Low Temp. Phys.} {\bf 167},
  707--712  (2012).

\bibitem{mates_2008}
J.~Mates, G.~C. Hilton, K.~D. Irwin, {\em et~al.}, ``Demonstration of a
  multiplexer of dissipationless superconducting quantum interference
  devices,'' {\em Applied Physics Letters} {\bf 92}(2), 023514  (2008).

\bibitem{hickish_etal_2016}
J.~{Hickish}, Z.~{Abdurashidova}, Z.~{Ali}, {\em et~al.}, ``{A Decade of
  Developing Radio-Astronomy Instrumentation using CASPER Open-Source
  Technology},'' {\em J. Astron. Instrum.} {\bf 5}, 1641001--12  (2016).

\bibitem{agostinelli_etal_2003}
S.~Agostinelli, J.~Allison, K.~Amako, {\em et~al.}, ``{Geant4 -- a simulation
  toolkit},'' {\em Nucl. Instrum. Meth. Phys. Res. A} {\bf 506}(3), 250 -- 303
  (2003).

\bibitem{allison_etal_2006}
J.~{Allison}, K.~{Amako}, J.~{Apostolakis}, {\em et~al.}, ``{Geant4
  developments and applications},'' {\em IEEE Trans. Nucl. Sci.} {\bf 53},
  270--278  (2006).

\bibitem{allison_etal_2016}
J.~Allison, K.~Amako, J.~Apostolakis, {\em et~al.}, ``{Recent developments in
  Geant4},'' {\em Nucl. Instrum. Meth. Phys. Res. A} {\bf 835}, 186 -- 225
  (2016).

\bibitem{spiga_etal_2004}
D.~{Spiga}, G.~{Pareschi}, O.~{Citterio}, {\em et~al.}, ``{Development of
  multilayer coatings (Ni/C-Pt/C) for hard x-ray telescopes by e-beam
  evaporation with ion assistance},'' in {\em UV and Gamma-Ray Space Telescope
  Systems},  G.~{Hasinger} and M.~J.~L. {Turner}, Eds., {\em Society of
  Photo-Optical Instrumentation Engineers (SPIE) Conference Series} {\bf 5488},
  813--819  (2004).

\bibitem{madsen_etal_2018}
K.~K. {Madsen}, F.~{Harrison}, D.~{Broadway}, {\em et~al.}, ``{Optical
  instrument design of the high-energy x-ray probe (HEX-P)},'' in {\em Space
  Telescopes and Instrumentation 2018: Ultraviolet to Gamma Ray},  J.-W.~A.
  {den Herder}, S.~{Nikzad}, and K.~{Nakazawa}, Eds., {\em Society of
  Photo-Optical Instrumentation Engineers (SPIE) Conference Series} {\bf
  10699}, 106996M  (2018).

\bibitem{gudmundsson_etal_2015}
J.~E. {Gudmundsson}, P.~A.~R. {Ade}, M.~{Amiri}, {\em et~al.}, ``{The thermal
  design, characterization, and performance of the SPIDER long-duration balloon
  cryostat},'' {\em Cryogenics} {\bf 72}, 65--76  (2015).

\bibitem{EBEX_cryo}
B.~{Reichborn-Kjennerud}, A.~M. {Aboobaker}, P.~{Ade}, {\em et~al.}, ``{EBEX: a
  balloon-borne CMB polarization experiment},'' in {\em Millimeter,
  Submillimeter, and Far-Infrared Detectors and Instrumentation for Astronomy
  V},  {\em Society of Photo-Optical Instrumentation Engineers (SPIE)
  Conference Series} {\bf 7741}, 77411C  (2010).

\bibitem{blast_cryo}
L.~M. {Fissel}, P.~A.~R. {Ade}, F.~E. {Angil{\`e}}, {\em et~al.}, ``{The
  balloon-borne large-aperture submillimeter telescope for polarimetry:
  BLAST-Pol},'' in {\em Millimeter, Submillimeter, and Far-Infrared Detectors
  and Instrumentation for Astronomy V},  {\em Society of Photo-Optical
  Instrumentation Engineers (SPIE) Conference Series} {\bf 7741}, 77410E
  (2010).

\bibitem{boomerang_crill}
B.~P. {Crill}, P.~A.~R. {Ade}, D.~R. {Artusa}, {\em et~al.}, ``{BOOMERANG: A
  Balloon-borne Millimeter-Wave Telescope and Total Power Receiver for Mapping
  Anisotropy in the Cosmic Microwave Background},'' {\em The Astrophysical
  Journal Supplement Series} {\bf 148}, 527--541  (2003).

\bibitem{ADR_FAA_ref}
G.~W. {Wilson} and P.~T. {Timbie}, ``{Construction techniques for adiabatic
  demagnetization refrigerators using ferric ammonium alum},'' {\em Cryogenics}
  {\bf 39}, 319--322  (1999).

\bibitem{chase_dilutor}
S.~T. {Chase}, T.~L.~R. {Brien}, S.~M. {Doyle}, {\em et~al.}, ``{Pre-cooling a
  $^{3}$He/$^{4}$He dilutor module with a sealed closed-cycle continuous
  cooler},'' in {\em Materials Science and Engineering Conference Series},
  {\em Materials Science and Engineering Conference Series} {\bf 502}, 012134
  (2019).

\bibitem{kislat_etal_2017}
F.~{Kislat}, B.~{Beheshtipour}, P.~{Dowkontt}, {\em et~al.}, ``{Design of the
  Telescope Truss and Gondola for the Balloon-Borne X-ray Polarimeter
  X-Calibur},'' {\em J. Astron. Instrum.} {\bf 6}, 1740003  (2017).

\bibitem{abarr_etal_2020b}
Q.~{Abarr}, H.~{Awaki}, M.~G. {Baring}, {\em et~al.}, ``{XL-Calibur - a
  second-generation balloon-borne hard X-ray polarimetry mission},'' {\em
  Astropart. Phys.} {\bf 126}, 102529  (2021).

\bibitem{abarr_etal_2021}
Q.~{Abarr}, B.~{Beheshtipour}, M.~{Beilicke}, {\em et~al.}, ``{Performance of
  the X-Calibur hard X-ray polarimetry mission during its 2018/19 long-duration
  balloon flight},'' {\em Astropart. Physics} {\bf 143}, 102749  (2022).

\bibitem{matsumoto_etal_2016}
H.~{Matsumoto}, H.~{Awaki}, A.~{Furuzawa}, {\em et~al.}, ``{Ray-tracing
  simulation and in-orbit performance of the ASTRO-H hard x-ray telescope
  (HXT)},'' in {\em Space Telescopes and Instrumentation 2016: Ultraviolet to
  Gamma Ray},  J.-W.~A. {den Herder}, T.~{Takahashi}, and M.~{Bautz}, Eds.,
  {\em Society of Photo-Optical Instrumentation Engineers (SPIE) Conference
  Series} {\bf 9905}, 990541  (2016).

\bibitem{shirazi_etal_2022}
F.~{Shirazi}, M.~{Arman Hossen}, D.~{Becker}, {\em et~al.}, ``{The 511-CAM
  Mission: A Pointed 511 keV Gamma-Ray Telescope with a Focal Plane Detector
  Made of Stacked Transition Edge Sensor Microcalorimeter Arrays},'' {\em arXiv
  e-prints} , arXiv:2206.14652  (2022).

\bibitem{barlow_1987}
R.~{Barlow}, ``{Event Classification Using Weighting Methods},'' {\em J. Comp.
  Phys.} {\bf 72}, 202--219  (1987).

\bibitem{gaskin_etal_2015}
J.~A. {Gaskin}, M.~C. {Weisskopf}, A.~{Vikhlinin}, {\em et~al.}, ``{The X-ray
  Surveyor Mission: a concept study},'' in {\em UV, X-Ray, and Gamma-Ray Space
  Instrumentation for Astronomy XIX},  {\em Proc. SPIE} {\bf 9601}, 96010J
  (2015).

\bibitem{bandler_etal_2016}
S.~R. {Bandler}, J.~S. {Adams}, J.~A. {Chervenak}, {\em et~al.}, ``{Development
  of x-ray microcalorimeter imaging spectrometers for the X-ray Surveyor
  mission concept},'' in {\em Space Telescopes and Instrumentation 2016:
  Ultraviolet to Gamma Ray},  {\em Proc. SPIE} {\bf 9905}, 99050Q  (2016).

\bibitem{lynx}
{Physics of the Cosmos Program Office}, ``{Program Annual Technology Report},''
   (2017).

\end{thebibliography}
\bibliographystyle{spiejour}   

\end{document}